\documentclass{article}

\usepackage{arxiv}

\usepackage[utf8]{inputenc} 
\usepackage[T1]{fontenc}    
\usepackage{hyperref}       
\usepackage{url}            
\usepackage{booktabs}       
\usepackage{amsfonts}       
\usepackage{nicefrac}       
\usepackage{microtype}      
\usepackage{natbib}
\usepackage{adjustbox}
\usepackage{verbatim}
\usepackage{multirow}
\usepackage{amsmath,amssymb,amsfonts,pifont}
\usepackage{mathtools}
 \usepackage{hyperref}
\usepackage{ulem}
\usepackage{soul}
\usepackage{blkarray, bigstrut}
\usepackage{multirow}
\usepackage{booktabs}
\usepackage{subfigure}
\usepackage{amsthm}
\usepackage{lineno}
\usepackage{tikz}
\usepackage{amsmath}
\usepackage{graphicx}
\usepackage{enumitem}
\usepackage{subcaption}
\usepackage{colortbl}
\usetikzlibrary{positioning, calc, shapes.geometric, arrows.meta}
\usepackage{pgf}
\usepackage{cleveref}
\usepackage{centernot}
\usepackage{wrapfig}
\usepackage{changepage}
\usepackage{algorithm}
\usepackage{algpseudocode}
\floatname{algorithm}{Pseudocode}
\usepackage{xcolor}
\definecolor{red}{rgb}{0,0,0}

\title{Optimizing Treatment Allocation \\in the Presence of Interference}

\date{September 30, 2024}	

\author{Daan Caljon\thanks{Corresponding author: \texttt{daan.caljon@kuleuven.be}.}\\
	KU Leuven\\
	\And
	Jente Van Belle\\
	KU Leuven\\
	\And
        Jeroen Berrevoets\\
        University of Cambridge\\
	\And
        Wouter Verbeke\\
        KU Leuven\\
}

\renewcommand{\shorttitle}{Optimizing Treatment Allocation in the Presence of Interference}

\hypersetup{
pdftitle={Using dynamic loss weighting to boost improvements in forecast stability},
pdfsubject={cs.LG, stat.ML, stat.ME},
pdfauthor={Daan Caljon, Jeff Vercauteren, Simon De Vos, Wouter Verbeke, Jente Van Belle},
pdfkeywords={Deep learning, Dynamic hyperparameter tuning, Rolling origin forecast instability, Global models, N-BEATS},
}

\begin{document}
\maketitle

\begin{abstract}
In Influence Maximization (IM), the objective is to---given a budget---select the optimal set of entities in a network to target with a treatment so as to maximize the total effect. For instance, in marketing, the objective is to target the set of customers that maximizes the total response rate, resulting from both direct treatment effects on targeted customers and indirect, spillover, effects that follow from targeting these customers. Recently, new methods to estimate treatment effects in the presence of network interference have been proposed. However, the issue of how to leverage these models to make better treatment allocation decisions has been largely overlooked. Traditionally, in Uplift Modeling (UM), entities are ranked according to estimated treatment effect, and the top entities are allocated treatment. Since, in a network context, entities influence each other, the UM ranking approach will be suboptimal. The problem of finding the optimal treatment allocation in a network setting is \textcolor{red}{NP-hard,} and generally has to be solved heuristically. To fill the gap between IM and UM, we propose OTAPI: Optimizing Treatment Allocation in the Presence of Interference to find solutions to the IM problem using treatment effect estimates. OTAPI consists of two steps. First, a causal estimator is trained to predict treatment effects in a network setting. Second, this estimator is leveraged to identify an optimal treatment allocation by integrating it into classic IM algorithms. We demonstrate that this novel method outperforms classic IM and UM approaches on both synthetic and semi-synthetic datasets. 
\end{abstract}


\section{Introduction}
A well-studied and important problem within network analysis is Influence Maximization (IM) \citep{kempe2003maximizing,domingos2001mining,richardson2002mining,azaouzi2021new}. 
The primary goal in IM is to identify the set of entities in a network to target with a particular intervention, e.g., a vaccine or a marketing campaign, in order to maximize the intervention's effect across the entire network given a budget $k$. 
The effect of treating one entity can cascade through the network, affecting connected entities as well. Consequently, careful analysis is essential to allocate treatment efficiently. For instance, in viral marketing, (potential) customers influence each other's purchasing decisions \citep{goldenberg2001talk}.  Hence, when determining whom to target with a marketing campaign, merely estimating the Individual Treatment Effects (ITEs), as commonly done in Uplift Modeling (UM) \citep{devriendt2018literature}, may not suffice. Considering interference effects---where treating one entity may influence others---is critical to identify the optimal set of entities to target. Practical applications of this problem include, e.g., vaccination and marketing campaigns. 

\textbf{Influence Maximization.}\; 
The IM problem can be formalized as follows: given a network $\mathcal{G} =(\mathcal{N}, \mathcal{E})$, where $\mathcal{N}$ represents the set of nodes and $\mathcal{E}$ denotes the set of edges, 
 the objective is to identify the set of $k$ seed nodes that are to be treated in order to maximize the total influence spread, i.e., the total effect of treatment allocation. 
 \textcolor{red}{This problem has been shown to be NP-hard \citep{kempe2003maximizing}.}
 In the IM literature, the influence spread of a certain treatment allocation is typically modeled using a diffusion process. This process describes how information or behavior disseminates through the network \citep{granovetter1978threshold,goldenberg2001talk}. 

We identify three significant limitations in the current approaches to address the IM problem. First, \textit{node features are mostly ignored}. Nevertheless, these features may contain important information regarding the spread of behaviors or effects through the network \citep{ogburn2014causal,ma2021causal}. Second, \textit{existing methods typically assume a homogeneous \textcolor{red}{direct} effect of treatment}. More specifically, it is assumed that all entities will respond positively (e.g., purchase the product) if they receive treatment. Yet, in many applications, treatment effects vary among entities, and the susceptibility to treatment often depends on entity characteristics \citep{shalit2017estimating,athey2016recursive}. Last, the most commonly used \textit{diffusion processes rely on assumptions rather than data}. However, employing an ill-defined diffusion process to solve the IM problem may lead to suboptimal outcomes \citep{aral2018social}. 

\textbf{Uplift Modeling.}\;
Another method for optimizing treatment allocation is UM \citep{devriendt2018literature}. This approach involves estimating the ITE for each entity, ranking them according to these estimates, and selecting the top $k$ ranked entities with the highest expected ITE for treatment. An important limitation of this approach, however, is its \textit{assumption that entities are independent of each other}. This assumption is violated in network settings where entities influence each other, such as marketing or vaccination campaigns. Recently, advanced causal inference methods have been proposed to cope with such network effects, known as spillover effects, by incorporating network information to estimate treatment effects \citep{forastiere2021identification,jiang2022estimating,cai2023generalization,huang2023modeling}. However, even with the capability to estimate treatment effects in the presence of interference, determining the optimal treatment allocation has been largely overlooked. Ranking entities according to their estimated treatment effects is no longer a possibility because the decision to treat one entity may impact the decision whether or not to treat another entity. 
\begin{figure}[t]
  \centering
  \includegraphics[width=1.05\linewidth]{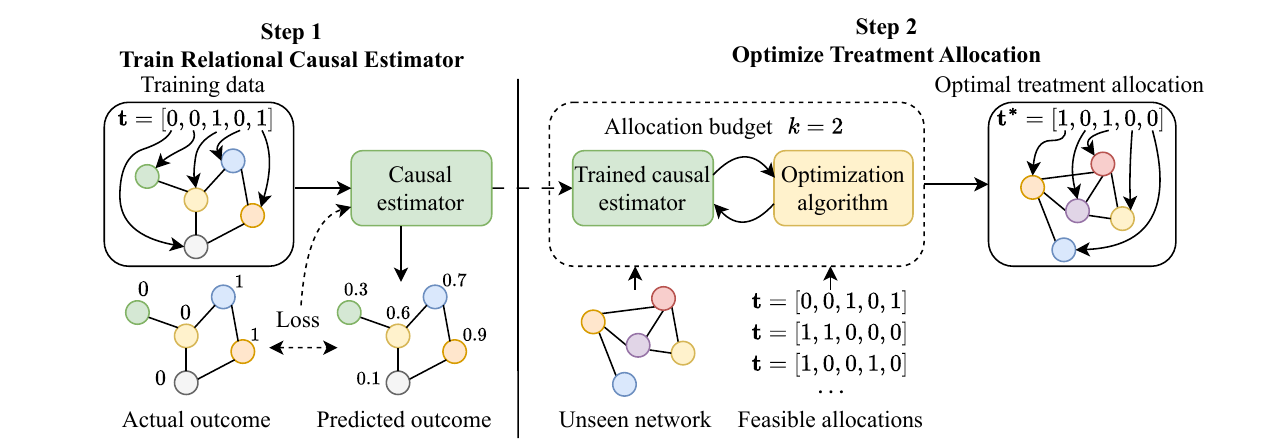}
  \caption{\textbf{Overview of OTAPI.} In \textbf{Step 1}, a relational causal estimator is trained on historical data. It receives the treatment assignment $\mathbf{t}$ and the network $\mathcal{G}$, including the features $\mathbf{x}_i$ for each node $i$ (represented by the colors) as input. The estimator outputs the predicted outcomes $\hat{y}_i$ for each node. To train the model, these predictions are compared with the actual outcomes $y_i$. In \textbf{Step 2}, an optimization algorithm \textcolor{red}{(e.g., a genetic or greedy algorithm)} is used to find the optimal treatment allocation $\mathbf{t^*}$ for a potentially unseen network (e.g., a network that has changed over time). This optimization algorithm takes as input the new network and the set of feasible allocations, which depends on the budget $k$. The trained causal model is leveraged to estimate the total effect of treatment allocation, which is used as the objective function in the optimization algorithm.}
  \label{fig:framework}
  \rule{\linewidth}{.5pt}
  \vspace{-15pt}
\end{figure}

\textbf{OTAPI.}\; 
We propose OTAPI: Optimizing Treatment Allocation in the Presence of Interference, a novel method that combines the strengths of both UM and IM approaches 
to identify improved, data-driven, solutions to the IM problem. Our method consists of two steps. First, a relational causal estimator is trained using observational data, e.g., from historical marketing or vaccination campaigns. A relational causal estimator is an estimator that can leverage both individual and network information to estimate causal effects.
This estimator allows estimating the total effect of any treatment allocation and controls for confounding. 
\textcolor{red}{Second, the treatment effect estimates from this causal estimator are 
integrated
into existing IM algorithms
in order to optimize the treatment allocation.}
An overview of OTAPI is depicted in \Cref{fig:framework}.

\textbf{Contributions.}\; \textcolor{red}{The challenge of data-driven treatment allocation that accounts for interference} is yet to be addressed. We conjecture that leveraging relational causal estimators will aid in finding the optimal treatment allocation. To this end, \textbf{(1)} we formalize the problem of finding the optimal treatment allocation in the presence of interference; \textbf{(2)} we present OTAPI, a new two-step method for optimizing treatment allocation in the presence of interference; and \textbf{(3)} we run extensive experiments on both synthetic and semi-synthetic data to empirically demonstrate that OTAPI finds better treatment allocations (in terms of the total effect of treatment allocation) than traditional methods used in UM and IM.

This article is structured as follows. In the following section, a comprehensive review of treatment effect estimation in the presence of interference and the IM problem is provided. Next, the problem setup and methodology are presented. Subsequently, we describe the experimental setup and analyze the results, before concluding this article and discussing the limitations and directions for future work.

\section{Related work}

\subsection{Treatment effect estimation in the presence of interference}
A substantial body of literature has focused on treatment effect estimation when entities are independent \citep{johansson2016learning,yao2018representation,devriendt2018literature}. Within this literature, an important and extensively studied problem is 
confounding bias, \textcolor{red}{also sometimes referred to as the covariate shift problem} \citep{rosenbaum1983central,shalit2017estimating,vairetti2024propensity}. This bias is associated with the treatment assignment mechanism, whereby treatment is not assigned randomly. Consequently, the distributions of the control and treatment groups will differ. In most applications, some form of mechanism or policy typically governs treatment assignment, leading to potential bias in treatment effect estimates. For instance, in marketing, a company might target a specific subgroup of the population. As it may be costly or even impossible to randomize the treatment assignment, (potentially biased) observational data has to be used. 

Recent approaches construct balanced representations of the distributions of the control and treatment groups to mitigate confounding bias \citep{shalit2017estimating,johansson2016learning,yao2018representation,hassanpour2019learning}. In a network setting, where entities can influence each other, mitigating confounding bias becomes more challenging. The features and treatments of connected entities might influence the probability of receiving treatment, introducing confounding bias as well \citep{ma2021causal,jiang2022estimating}. 

In estimating treatment effects in the presence of interference, a common approach involves defining an exposure mapping \citep{aronow2017estimating}, which generates a variable (or set of variables) based on the treatments of other relevant entities. 
Many methods assume this exposure mapping to be the percentage of directly connected entities, known as neighbors, that received treatment, and use Graph Neural Networks (GNNs) \citep{kipf2016semi,xu2018powerful} to learn a model from observational data \citep{ma2021causal,jiang2022estimating,cai2023generalization}. Representation balancing techniques are employed to mitigate confounding bias. Some recent works propose to relax the assumption on the exposure mapping to allow for heterogeneous peer influence, where the spillover effect may vary depending on the features of connected entities \textcolor{red}{\citep{huang2023modeling,adhikari2023inferring,zhao2024learning}}.  
Another line of work focuses on leveraging network information, such as features of neighbors, to better control for hidden confounders. For example, a person's community affiliations may help to represent their socioeconomic status. However, these methods typically ignore spillover effects, assuming that the treatment of one entity does not causally affect other entities \citep{guo2020learning,guo2021ignite}. \

\subsection{Influence Maximization problem}
The IM problem was initially introduced by \citet{domingos2001mining}, and later formalized as a combinatorial optimization problem by \citet{kempe2003maximizing}. 
A multitude of methods to tackle the IM problem rely on diffusion processes, which define how influence propagates through the network. A wide variety of diffusion processes have been proposed, each with its distinct assumptions about the dissemination of behavior or information \citep{goldenberg2001talk,granovetter1978threshold,kempe2003maximizing,ye2022influence}. To identify the optimal allocation set, existing methods mostly employ a greedy algorithm proposed by \citet{kempe2003maximizing}. This algorithm uses Monte Carlo (MC) simulations based on the diffusion process to estimate the expected influence spread, i.e., the expected number of activated entities. However, its computational complexity may render it infeasible in terms of runtime, especially when networks become large. To this end, several methods have been developed to reduce the number of MC simulations, such as CELF \citep{leskovec2007cost} and CELF++ \citep{goyal2011celf++}. Other methods were developed to partition the network into communities, significantly reducing the runtime of an MC simulation \citep{wang2010community,bozorgi2017community}. Alternatively, some approaches ignore diffusion processes altogether, instead solely using the network structure. These methods employ, for instance, the degree centrality, which is defined as the number of connections of a node. IM algorithms based on this measure include the degree \citep{kempe2003maximizing} and single discount heuristics \citep{chen2009efficient}.

All the aforementioned methods rely on the network structure or make assumptions regarding the diffusion process in order to arrive at a solution. It has been shown, however, that incorporating data to estimate the spread of influence through the network yields more reliable solutions \citep{aral2018social}. In the IM literature, considerable work has been done to employ propagation traces to estimate the parameters of specific diffusion processes \citep{goyal2010learning,saito2008prediction,bourigault2016representation,tran2022heterogeneous}. Propagation traces are data structures containing information about which entities were activated and when. \citet{goyal2011data} uses these traces to create a model that does not make any parametric assumptions about the diffusion process, but instead learns the expected influence spread directly from data. These methods, however, do not consider node features. The only exception is \citet{tran2022heterogeneous}, in which node features are leveraged to estimate the parameters in the Linear Threshold model, a commonly used diffusion process. In contrast to these approaches, our proposed method leverages node features, does not make parametric assumptions regarding the diffusion process, and does not use propagation traces. Moreover, OTAPI allows for \textit{out-of-sample} IM, as the network for which we try to find the optimal treatment allocation may be different from the network that was used for training the model.
This is an important advantage of OTAPI, as networks---especially social networks---typically change over time. \citet{ma2021causal} proposes a utility maximization approach for instance-level treatment assignment in the presence of interference using a policy network that leverages a relational causal estimator. The budget constraint is satisfied by integrating it into the loss function of the policy network and allowing some tolerance for violating it. In comparison, OTAPI directly optimizes treatment allocation for the network as a whole by leveraging algorithms that can support hard constraints.

\section{Problem setup}\label{SEC: Problem setup}
\textbf{Notation.}\; Consider an undirected network $\mathcal{G}=(\mathcal{N},\mathcal{E})$ where $\mathcal{N}$ represents the set of nodes and $\mathcal{E}$ denotes the set of edges between them. We define the set of neighbors of node $i$ as $\mathcal{N}_i$, which comprises nodes $j \in \mathcal{N}$ for which an edge \textcolor{red}{$\{i,j\}$ exists. In this context, $\{i, j\} = \{j, i\}$.} Additionally, we denote the set of nodes not in the neighborhood of $i$ as $\mathcal{N}_{-i}=\mathcal{N} \setminus (\mathcal{N}_i \cup \{i\})$.
Each node $i$ possesses a feature vector $\mathbf{X}_i \in \mathcal{X} \subseteq \mathbb{R}^d$. We will use nodes and entities interchangeably, depending on the context. 

Let $Y_i \in \mathcal{Y}$ represent the outcome of entity $i$. Without loss of generality, $\mathcal{Y} = \{0,1\}$, e.g., buying or not buying the product. We define $T_i \in \mathcal{T} = \{0,1\}$ as the treatment given to entity $i$. The treatments of $i$'s neighbors $\mathcal{N}_i$ is given by $\mathbf{T}_{\mathcal{N}_i} = \{T_j\}_{j \in \mathcal{N}_i}$. Following \citet{forastiere2021identification}, \textcolor{red}{\citet{jiang2022estimating}}, and \citet{aronow2017estimating}, an exposure mapping, denoted by 
$Z:\mathbf{T}_{\mathcal{N}_i} \to Z_i$, is defined to summarize $\mathbf{T}_{\mathcal{N}_i}$. This function describes how the outcome of an entity is dependent on $\mathbf{T}_{\mathcal{N}_i}$.  The domain $\mathcal{Z}$ of the exposure mapping $Z_i$ depends on the exact definition of the function $Z$ \citep{forastiere2021identification}. \textcolor{red}{We follow the potential outcomes framework \citep{rubin1974estimating}, where $Y_i(T_i=t,Z_i=z)$ denotes the outcome value entity $i$ would take if it receives treatment $t$ with a neighborhood exposure $z$. A potential outcome is not necessarily observed. For readability, we will sometimes write $\mathbb{E}\left[Y_i(T_i=t,Z_i=z)\mid \mathbf{X}_i=\mathbf{x}_i,\mathbf{X}_{\mathcal{N}_i}=\mathbf{x}_{\mathcal{N}_i}\right]$ as $\mathbb{E}\left[Y_i(t,z)\mid \mathbf{x}_i,\mathbf{x}_{\mathcal{N}_i}\right]$.} 

\begin{figure}
    \centering

        {\begin{tikzpicture}[>=stealth, node distance=2cm, on grid, auto,
        Xnode/.style={circle,draw,fill = cyan!20},
        Tnode/.style={circle,draw, fill = green!20},
        Ynode/.style={circle,draw, fill = yellow!20},
        every node/.style={minimum size=1cm}]
            \node[Xnode] (Xi) {$\mathbf{X}_i$};
        \node[Xnode, below=of Xi] (Xj) {$\mathbf{X}_j$};
        
        \node[Tnode, right=of Xi] (Ti) {$T_i$};
        \node[Tnode, below=of Ti] (Tj) {$T_j$};
        
        \node[Ynode, right=of Ti] (Yi) {$Y_i$};
        \node[Ynode, below=of Yi] (Yj) {$Y_j$};

            \path[->] (Xi) edge (Ti)
                     (Xi) edge (Tj)
                     (Xj) edge (Ti)
                     (Xj) edge (Tj)
                     (Xi) edge[bend left] (Yi)
                     (Xi) edge (Yj)
                     (Xj) edge (Yi)
                     (Xj) edge[bend right] (Yj)
                     (Ti) edge (Yi)
                     (Ti) edge (Yj)
                     (Tj) edge (Yi)
                     (Tj) edge (Yj);
        \end{tikzpicture}}
        \caption{{\bf Graphical representation of the causal structure.} Only two entities 
        are included
        for simplicity. $Y_i$ \textcolor{red}{might be} influenced by its own features $\mathbf{X}_i$, the features of its neighbors $\mathbf{X}_j$, its own treatment $T_i$, and the treatment of its neighbors $T_j$. Moreover, $T_i$ \textcolor{red}{might be} influenced by both $\mathbf{X}_i$ and $\mathbf{X}_j$.}
        \label{fig:Direct interference graph}

\end{figure}

\textbf{Assumptions.}\;
We assume that an entity's own treatment $T_i$, the treatments of its neighbors $\mathbf{T}_{\mathcal{N}_i}$, an entity's features $\mathbf{X_i}$ and the features of its neighbors $\mathbf{X}_{\mathcal{N}_i} = \{\mathbf{X}_j\}_{j \in \mathcal{N}_i}$ \textcolor{red}{might} causally influence its outcome  $Y_i$. The assumed causal structure is depicted in 
Figure \ref{fig:Direct interference graph} \textcolor{red}{as a Directed Acyclic Graph (DAG).} \textcolor{red}{While it relaxes the ``no interference" assumption, the DAG does not imply that we assume there always is interference. The presence of a causal arrow in this DAG implies that we do not assume anything about whether there truly is a causal effect \citep{pearl2009causality}.} Note that we do not consider \textit{contagion} effects, meaning the outcomes of different entities do not causally affect each other \citep{ogburn2014causal}. To summarize $\mathbf{T}_{\mathcal{N}_i}$ we use the mapping $Z_i = \frac{\sum_{j \in \mathcal{N}_i} T_j}{|\mathcal{N}_i|}$ \citep{forastiere2021identification,jiang2022estimating}\footnote{Note that more complex exposure mappings (e.g., multi-dimensional, based on node features, etc.) can also be defined/learned \citep{cai2023generalization,adhikari2023inferring,aronow2017estimating}.}. We assume that the decision to treat is taken simultaneously for all entities. Consequently, both $Z_i$ and $T_i$ are part of the intervention on $i$.

Traditionally, each entity has two potential outcomes: $Y_i(T_i=0)$ and $Y_i(T_i=1)$. However, in the presence of interference, the traditional consistency assumption, which states that there are no multiple versions of treatment, is violated. Here, depending on the treatment of $i$'s neighbors, $i$'s potential outcomes might differ. To address this, the consistency assumption
is relaxed by employing the exposure mapping $Z_i$ \citep{forastiere2021identification,jiang2022estimating}:

\textit{Consistency.} If $T_i = t$ and $Z_i=z$, then $Y_i = Y_i(t,z)$.


\noindent
Moreover, we \textcolor{red}{adopt} the overlap and strong ignorability assumptions in the presence of interference \citep{forastiere2021identification,jiang2022estimating}:

\textit{Overlap and strong ignorability.} $\exists \ \delta \in (0,1)$ such that $\delta < p(T_i=1|\mathbf{X}_i=\mathbf{x}_i,\mathbf{X}_{\mathcal{N}_i}=\mathbf{x}_{\mathcal{N}_i}) < 1-\delta$ and $Y_i(T_i=t,Z_i=z) \perp\!\!\!\perp
T_i,Z_i
\mid\mathbf{X}_i,\mathbf{X}_{\mathcal{N}_i}, \forall t \in \mathcal{T}, z\in \mathcal{Z}, \mathbf{X}_i \in \mathcal{X}$ and $ \mathbf{X}_{\mathcal{N}_i} = \{\mathbf{X}_j\}_{j\in \mathcal{N}_i}$. However, the traditional ignorability assumption no longer holds: $Y_i(T_i=t) \centernot{\perp\!\!\!\perp}$ $T_i \mid \mathbf{X}_i$ $\forall t \in \mathcal{T}$ and $\mathbf{X}_i \in \mathcal{X}$. \textcolor{red}{Under the stated assumptions, the causal estimands defined in the next paragraph are identifiable \citep{forastiere2021identification}, meaning that they can be expressed as functions of an observed data distribution. Consequently, these causal effects can be estimated from data. A proof of identifiability is provided in \Cref{app:identifiability}.}

\textbf{Treatment effect definitions.}\; 
\textcolor{red}{We now define our causal estimands of interest. The} \textit{Marginal Individual Treatment Effect (MITE)} is defined as:
\begin{equation}\label{eq:mite}
\tau_i(z)=\mathbb{E}\bigl[Y_i(1,z)-Y_i(0,z)\mid\mathbf{x}_i, \mathbf{x}_{\mathcal{N}_i}\bigr].
\end{equation}
\Cref{eq:mite} shows that MITE is dependent on the value of $Z_i=z$. \textcolor{red}{The difference between MITE and the traditional ITE is clarified in \Cref{app:MITE v ITE}.}


The \textit{spillover effect} is defined as:
\begin{equation}
\delta_i(t,z) = \mathbb{E}\bigl[Y_i(t,z)-Y_i(t,0)\mid\mathbf{x}_i, \mathbf{x}_{\mathcal{N}_i}\bigr].\label{eq:spillover definition}
\end{equation}
The spillover effect is the effect of the treatments of $i$'s neighbors on $i$, and not the effect of the treatment of $i$ on its neighbors. Furthermore, the spillover effect is dependent on the value of $T_i=t$.

Finally, we can also define the \textit{Individual Total Treatment Effect (ITTE)} as:
\begin{equation}
\omega_i(t,z) = \mathbb{E}\bigl[Y_i(t,z)-Y_i(0,0)\mid\mathbf{x}_i, \mathbf{x}_{\mathcal{N}_i}\bigr].
\end{equation}
The ITTE quantifies the total effect of the full treatment allocation on an entity. It is noteworthy that the ITTE can be rewritten in terms of MITE and spillover effect: 
\begin{equation}
    \omega_i(t,z)=
    \begin{cases}
        \delta_i(0,z) & \text{if } t = 0,\\
        \tau_i(0)+\delta_i(1,z) = \delta_i(0,z)+\tau_i(z) & \text{if } t = 1.
    \end{cases}
    \label{eq:omeqa equivalence}
\end{equation}
\textcolor{red}{A proof of this equivalence is provided in \Cref{app:decomposition ITTE}.}

\textbf{Objective.}\; 
We aim to find an optimal treatment allocation $\mathbf{t^*}= \{t_i\}_{i \in \mathcal{N}}$ that maximizes the \textit{Total Treatment Effect (TTE)}. The TTE is defined as the sum of ITTE $\sum_{i\in \mathcal{N}}
 \omega_i(t_i,z_i)$ and quantifies the total effect of treatment allocation on all entities. 
The problem can now be formalized as:
\begin{equation}
 \mathbf{t^*} = \underset{\mathbf{t}}{\arg\max}\sum_{i\in \mathcal{N}}
 \omega_i(t_i,z_i) \quad \text{s.t.} \quad \sum_{i\in \mathcal{N}} t_i \leq k. \label{eq:objective}
\end{equation}
%
Both $t_i$ and $z_i$ are decision variables and can be derived from $\mathbf{t}$. Now, if we have a model that can estimate the ITTEs for a given $\mathbf{t}$, we can use (the sum of) these estimates as our objective function to find an optimal treatment allocation $\mathbf{t^*}$.

\section{OTAPI}\label{sec:methodology}
To clearly position OTAPI, we first compare it to two alternative treatment allocation approaches: Uplift Modeling (UM) and classic Influence Maximization (IM). 

In \textit{UM}, the focus lies on finding the entities with the highest expected ITE. The goal in UM is essentially the same as in \Cref{eq:objective}. To maximize TTE, entities are ranked according to estimated ITE, and the top $k$ entities are given treatment. Alternatively, integrated approaches exist that rank entities directly without estimating ITE first \citep{devriendt2020learning}. Nevertheless, these approaches are not tailored to scenarios involving network interference because they rely solely on individual information. When network interference plays a significant role, ITE estimates may be biased, potentially resulting in suboptimal treatment allocation, where entities with substantial network effects might be overlooked and connected entities might all be treated while treating just one of them might have a similar effect in terms of TTE \citep{forastiere2021identification}. Moreover, each treatment allocation vector may result in a different ITTE in the presence of interference. Therefore, simply ranking entities according to ITTE no longer works because an entity's ITTE depends on the set of other entities that receive treatment.

\textit{Classic IM} methods focus on modeling network effects, often using diffusion processes to simulate interference dynamics. These require computationally expensive MC simulations to estimate TTE, also defined as influence spread in the IM literature. \textcolor{red}{Because it is often infeasible to estimate the TTE for every possible combination of treatment allocations in larger graphs, heuristics are used to limit the number of treatment allocation combinations for which TTE is estimated \citep{kempe2003maximizing,bucur2016influence}}. These TTE estimates lack any form of individualized treatment effect estimation and are commonly based on assumptions regarding the applicable diffusion process. 

\textit{OTAPI} addresses these limitations by combining the strengths of these two methods. It consists of two components: (i) a relational causal estimator that considers both individualized susceptibility to treatment and interference to estimate ITTE (\Cref{sec:otapi:ite}), and (ii) an allocation algorithm that uses the predictions made by the trained causal model 
to allocate treatments optimally (\Cref{sec:otapi:opt}). \textcolor{red}{A high-level pseudocode description of OTAPI is provided in \Cref{app:pseudo}.}



\subsection{Estimating treatment effects in the presence of interference} \label{sec:otapi:ite}

In order to optimally allocate treatment, accurate estimates of the ITTE are required.
To estimate unbiased treatment effects, we employ causal machine learning. OTAPI is agnostic to the estimator employed, but its performance directly depends on the accuracy of the estimates.

Given the assumptions in \Cref{SEC: Problem setup}, we use a causal estimator $\mathcal{M}(\mathbf{x}_i,\mathbf{x}_{\mathcal{N}_i},t_i,z_i)$ which provides unbiased estimates for the potential outcomes $\mathbb{E}\left[Y_i(T_i=t_i,Z_i=z_i)|\mathbf{x}_i,\mathbf{x}_{\mathcal{N}_i}\right]$. 
Since an entity can have an arbitrary number of neighbors, a GNN is employed to learn an aggregation of the neighbors' features. \citet{jiang2022estimating} shows that if the distribution $p(T_i|\mathbf{x}_i,\mathbf{x}_{\mathcal{N}_i})$, \textcolor{red}{i.e.,} the classic propensity score, \textcolor{red}{or} $p(Z_i|\mathbf{x}_i,\mathbf{x}_{\mathcal{N}_i},t_i)$, \textcolor{red}{i.e.,} the peer exposure score, \textcolor{red}{is biased (or when both are biased)}, the estimates for the treatment effects will also be biased. Given that treatments are typically not allocated randomly in observational data due to confounding, controlling for these distributions is necessary. 

To address the challenge of confounding bias in the presence of interference, representation balancing techniques can be applied \citep{ma2021causal,jiang2022estimating,shalit2017estimating}. This leads to more accurate estimates of the ITTE. In the experiments presented in \Cref{sec:experiments}, we adopt NetEst \citep{jiang2022estimating}, a GNN-based causal estimator that employs an adversarial approach to balance representations. Nevertheless, alternative relational causal estimators could also be employed. 

\subsection{Allocating treatments optimally} \label{sec:otapi:opt}

Assuming we can compute unbiased ITTE estimates for all entities as $\hat{\omega}_i(t_i,z_i)=\mathcal{M}(\mathbf{x}_i,\mathbf{x}_{\mathcal{N}_i},t_i,z_i)$ $-$ $ \mathcal{M}(\mathbf{x}_i,\mathbf{x}_{\mathcal{N}_i},0,0)$, we can plug these estimates into the objective (\Cref{eq:objective}). 
As already explained, the strategy of ranking entities according to estimated ITTE no longer works. To address this, we rely on heuristics developed in the IM literature and plug data-driven estimates of TTE into these algorithms. Generally, these heuristics maximize TTE by iteratively improving a (set of) solution(s). Some examples include a greedy algorithm \citep{kempe2003maximizing}, a genetic algorithm \citep{bucur2016influence}, and simulated annealing \citep{jiang2011simulated}. OTAPI can use any of these algorithms to optimize treatment allocation.


\section{Experiments}\label{sec:experiments}
\subsection{Experimental setup} \label{sec: experimental setup}
\subsubsection{Data}
To properly evaluate treatment allocation methods in the presence of interference, the ground truth TTE is required. However, due to the fundamental problem of causal inference \citep{holland1986statistics}, the TTE is unobserved. Therefore, in causal inference, (semi-)synthetic data is often used \citep[see, e.g.,][]{vanderschueren2023optimizing,berrevoets2020}. Hence, we generate synthetic and semi-synthetic data according to the causal diagram in \Cref{fig:Direct interference graph}. 

\textbf{Synthetic data.}\; For each synthetic dataset, 3 separate networks (train, validation, and test) are generated according to the Barabási-Albert random network model \citep{barabasi1999emergence}. Each network consists of 5,000 nodes. The hyperparameter which determines the number of edges attached from each new node is set to $m=2$.
The data-generating process (DGP) is inspired by \citet{jiang2022estimating}.

\textcolor{red}{For each node, a feature vector of $10$ features is randomly sampled according to a standard normal distribution:
${x}_i^j \sim \mathcal{N}(0, 1), j=1,\dots,10$.
We first generate the parameters that will be used later on in the DGP:
\begin{align*}
& \begin{aligned}
&w^{XT}_j \sim \text{Unif}(-1, 1) \quad \text{for } j \in \{1, 2, \dots, 10\}; \\
&w^{XY}_j \sim \text{Unif}(-1, 1) \quad \text{for } j \in \{1, 2, \dots, 10\}; \\
&w^{TY}_j \sim \text{Unif}(-1, 1) \quad \text{for } j \in \{1, 2, \dots, 10\}; \\
\end{aligned} &&
\begin{aligned}
&\mathbf{w}^{XT} = [w^{XT}_1, w^{XT}_2, \dots, w^{XT}_{10}], \\
&\mathbf{w}^{XY} = [w^{XY}_1, w^{XY}_2, \dots, w^{XY}_{10}], \\
&\mathbf{w}^{TY} = [w^{TY}_1, w^{TY}_2, \dots, w^{TY}_{10}], \\
\end{aligned}
\end{align*}}
where Unif is the uniform distribution. These parameters are used to quantify the effect of $\mathbf{X_i}$ on ${T_i}$, $\mathbf{X_i}$ on $Y_i$, and ${T_i}$ on $Y_i$, respectively. The treatment $t_i$ is generated as follows:
\begin{equation*}
     t_i \sim \text{Bernoulli}(p_{t_i}); \quad  p_{t_i} = \sigma(\beta_{XT} \cdot p_i + \beta_{{\mathcal{N}}T} \cdot p_{\mathcal{N}_i}),
\end{equation*}
\textcolor{red}{with $p_i = \mathbf{w}^{XT} \cdot \mathbf{x}_i$ and $p_{\mathcal{N}_i}= \frac{\sum_{j \in \mathcal{N}_i} p_j}{|\mathcal{N}_i|}$.}
Here, $\sigma$ is the sigmoid function.
Note that confounding is introduced by generating propensity scores $p_i$ and $p_{\mathcal{N}_i}$ based on $\mathbf{x}_i$ and $\mathbf{x}_{\mathcal{N}_i}$, respectively. 
Now, we describe how the outcome $y_i$ is generated:
\begin{equation*}
    \textcolor{red}{
    y^{prob}_i = \sigma(\beta_0 + h_i \cdot \beta_{{individual}} \cdot t_i + h_i \cdot \beta_{{spillover}} \cdot z_i + \beta_{XY} \cdot u_i + \beta_{\mathcal{N}Y} \cdot u_{\mathcal{N}_i} + \beta_{\epsilon} \cdot \epsilon); \quad \epsilon \sim \mathcal{N}(0,1),}
\end{equation*}
where $h_i = \sigma(\mathbf{w}^{TY} \cdot \mathbf{x}_i) + b_{TY}$, $u_i = \sigma(\mathbf{w}^{XY} \cdot \mathbf{x}_i)$, and $u_{\mathcal{N}_i} = \frac{\sum_{j \in \mathcal{N}_i} u_j}{|\mathcal{N}_i|}$.
Recall that $z_i$ is the ratio of $i$'s neighbors that received treatment $z_i = \frac{\sum_{j \in \mathcal{N}_i} t_j}{|\mathcal{N}_i|}$. \textcolor{red}{Finally, $y_i$ is sampled using a Bernoulli distribution $y_i \sim \text{Bernoulli}(y_i^{prob})$.}
$h_i$ is used to make the treatment effect heterogeneous, i.e., dependent on the features. Note that the individual effect is 100\% correlated with the spillover effect. This choice is motivated by the fact that if someone is more/less susceptible to treatment, they will most likely also be more/less susceptible to the treatment of neighbors. The following setting for the parameters was used for all experiments:
$\beta_0 = -3$, 
$\beta_{{individual}} = 1$, 
$\beta_{XT} = 1$, 
$\beta_{\mathcal{N}T} = 0.5$,
$\beta_{XY} = 0.7$, 
$\beta_{\mathcal{N}Y} = 0.2$, 
$\beta_{\epsilon} = 0.05$, 
and $b_{TY} = 4$.
The intuition behind these parameters is that the individual features and treatments will always be more impactful than those of neighbors. $\beta_{spillover}$ is varied between 0 and 0.7 across experiments.

\textbf{Semi-synthetic data.} \;\textcolor{red}{Following \citet{jiang2022estimating} and \citet{guo2020learning}, we use the BlogCatalog (BC) and Flickr datasets
to construct semi-synthetic data: the network structures are derived from the real-world data, while both the treatments and outcomes are simulated. 
Additionally, we use the Enron email interaction dataset \citep{network-repository} to generate another semi-synthetic dataset.}

\textcolor{red}{BC is a social network platform where people can share their blogs. In the BC dataset, the bloggers are the nodes, and the edges represent whether two bloggers are friends on the platform. The node features are bag-of-words vectors derived from the keywords in the bloggers' textual descriptions. Hence, the features are sparse and high-dimensional. Therefore, following \citet{guo2020learning}, we use Latent Dirichlet Allocation (LDA) \citep{blei2003latent} to reduce the dimensionality, and construct more informative features. Product sales are simulated as the outcome variable, influenced by a targeted marketing campaign that represents the treatment. Based on their features, some bloggers may be more susceptible to the campaign than others. Additionally, if bloggers start posting or talking about the product of interest, word-of-mouth effects may amplify its visibility and drive product sales.}

\textcolor{red}{Flickr is a social media platform where users can share images and videos. In the corresponding dataset, each node represents a user, and edges denote friendship links between users. Node features capture the tags associated with each user's content interests. As with the BC dataset, these features are represented as high-dimensional, sparse vectors. Hence, the same approach (LDA) is adopted to reduce the dimensionality. The downstream task is defined analogously to the BC case: the treatment is a marketing campaign intended to promote a product, and the outcome is product sales.}

\textcolor{red}{In the Enron e-mail interaction dataset, each node represents an employee, and edges correspond to email communications between them. Social ties among employees are a key factor in turnover \citep{soltis2013social}, which can be socially contagious---when one employee leaves, the likelihood that their close colleagues will also leave may increase \citep{porter2021turnover}. This highlights the importance of the social network in understanding turnover dynamics. Identifying employees with high network impact is therefore crucial, as retaining them---through, for example, salary increases, additional benefits, or professional development opportunities---can help prevent cascading departures and maintain organizational stability.}
\textcolor{red}{Since no node features are provided, we generate them using node2vec \citep{grover2016node2vec}, which produces feature embeddings based on the network structure. In this dataset, the treatment corresponds to offering a pay raise, and the outcome is whether the employee eventually leaves the organization (i.e., turnover).}

\textcolor{red}{We follow the dataset construction procedure described in \citet{jiang2022estimating}. To evaluate the out-of-sample performance of OTAPI, each network is partitioned into training, validation, and test sets using the METIS graph partitioning algorithm \citep{karypis1998fast}. For the Flickr and BC datasets, we use the same data splits as in \citet{jiang2022estimating}. The feature dimension of these two datasets is reduced to 10. To assess OTAPI's performance with more node features, the feature dimension for the Enron dataset is set to 100. Summary statistics for all semi-synthetic datasets are presented in \Cref{tab:datasets}. After feature generation, the same DGP as in the synthetic datasets is used. For the semi-synthetic datasets, we set $\beta_{spillover}=0.3$; all other parameters remain unchanged.}


\begin{table}[t]
    \centering
    \begin{tabular}{lcccccccccc} 
        \toprule 
        & \multicolumn{3}{c}{Flickr} & \multicolumn{3}{c}{BC} & \multicolumn{3}{c}{\textcolor{red}{Enron}} \\
        \cmidrule(lr){2-4} \cmidrule(lr){5-7} \cmidrule(lr){8-10}
        
        & Train & Validation & Test & Train & Validation & Test & Train & Validation & Test\\
        
        \midrule 
        
        Nodes & 2,482 & 2,461 & 2,358 & 1,716 & 1,696 & 1,784 & 11,213& 11,232 & 11,232\\
        Edges & 46,268 & 14,419 & 23,529 & 17,937 & 25,408 & 14,702 & 33,228 &55,046 &63,027 \\
        Features & 12,047 & 12,047 & 12,047 & 8,189 & 8,189 & 8,189 &100 & 100&100\\
        
        \bottomrule 
    \end{tabular}
    \caption{\textcolor{red}{Summary statistics for the Flickr, BC, and Enron datasets.}}
    \label{tab:datasets}
\end{table}








\begin{figure}[h]
    \centering
    \resizebox{0.45\textwidth}{!}{\begin{tikzpicture}[x=2cm, y=1.5cm, >=Stealth]

\tikzstyle{inputnode} = [circle, draw, minimum size=1cm, fill=cyan!20]
\tikzstyle{gnnnode} = [rectangle, draw, rounded corners, minimum size=1cm, fill=green!20]
\tikzstyle{hiddennode} = [rectangle, draw, rounded corners, minimum width=0.3cm, minimum height=1cm, fill=green!20]
\tikzstyle{outputnode} = [circle, draw, minimum size=1cm, fill=yellow!20]
\tikzstyle{discriminator} = [circle, draw=none, fill=none]
\node[inputnode] (xN) at (0,0) {$\mathbf{x}_{\mathcal{N}_i}$};
\node[inputnode, below=0.5cm of xN] (xi) {$\mathbf{x}_i$};

\node[gnnnode, right=0.5cm of xN] (gnn) {GCN};

\node[hiddennode, right=0.5cm of gnn] (hidden) {};


\node[outputnode, right=0.7cm of hidden] (phi) {$\mathbf{\phi}_i$};

\coordinate (phiup) at ($(phi.north) + (0,1cm)$);
\coordinate (phidown) at ($(phi.south) - (0,1cm)$);

\node[inputnode,right= 0.5cm of phiup] (T) {$t_i$};
\node[inputnode,right=0.5cm of phidown] (Z) {$z_i$};

\node[hiddennode, right= 1cm of T] (hiddenZ){};
\node[hiddennode, right= 1cm of Z] (hiddenT){};
\node[hiddennode, right= 2cm of phi] (hiddenY){};

\node[discriminator, above= -0.2cm of hidden] (encoder){$e_{\phi}$};
\node[discriminator, above= -0.2cm of hiddenT] (d_t){$d_T$};
\node[discriminator, above= -0.2cm of hiddenZ] (d_Z){$d_Z$};
\node[discriminator, above= -0.2cm of hiddenY] (p_Y){$p_Y$};

\node[outputnode, right=1cm of hiddenY] (Yhat){$\hat{y}_i$};
\node[outputnode, above=0.5cm of Yhat] (Zhat){$\hat{z}_i$};
\node[outputnode, below=0.5cm of Yhat] (That){$\hat{t}_i$};

\coordinate (phi_T_1) at ($(phi.north) + (0,1.8cm)$);
\coordinate (phi_T_above) at ($(T.north) + (0,0.3cm)$);
\coordinate (phi_T_east) at ($(T.east) + (0.3cm,0)$);
\coordinate (phi_T_east_above) at ($(phi_T_east.north) + (0,0.25cm)$);
\coordinate (dz_phi) at ($(hiddenZ.west) + (0,0.25cm)$);

\coordinate (phi_Z_1) at ($(phi.south) - (0,1.8cm)$);
\coordinate (phi_Z_below) at ($(Z.south) - (0,0.3cm)$);
\coordinate (phi_Z_east) at ($(Z.east) + (0.3cm,0)$);

\coordinate (phi_right) at ($(phi.east) + (0.5cm,0)$);

\coordinate (phi_right_Z) at ($(phi_right) - (0,0.25cm)$);
\coordinate (hiddenY_Z) at ($(hiddenY.west) - (0,0.25cm)$);
\coordinate (phi_right_T) at ($(phi_right) + (0,0.25cm)$);
\coordinate (hiddenY_T) at ($(hiddenY.west) + (0,0.25cm)$);


\draw[->] (xN) -- (gnn);
\draw[->] (xi) -| (hidden);
\draw[->] (gnn) -- (hidden);
\draw[->] (hidden) -- (phi);
\draw[->] (phi) -- (hiddenY);

\draw[->] (T) --(phi_right_T) -- (hiddenY_T);
\draw[->] (Z) --(phi_right_Z) -- (hiddenY_Z);
\draw[->] (phi) --(phi_T_1) -- (phi_T_above) -| (phi_T_east_above) -- (dz_phi);

\draw[->] (phi) --(phi_Z_1) -- (phi_Z_below) -| (phi_Z_east) -- (hiddenT);

\draw[->] (T) -- (hiddenZ);

\draw[->] (hiddenY) -- (Yhat);
\draw[->] (hiddenZ) -- (Zhat);
\draw[->] (hiddenT) -- (That);

\end{tikzpicture}}
    \caption{NetEst architecture.}
    \label{fig:netest_tikz}
\end{figure}

\subsubsection{Causal estimator}
As already mentioned in \Cref{sec:methodology}, we will use a modified version of \textit{NetEst} \citep{jiang2022estimating} to estimate ITTEs. We opt for NetEst because of its strong empirical performance, solid theoretical foundation, and the availability of its source code. \textcolor{red}{Note, however, that other relational causal estimators could also be used.} The architecture is shown in \Cref{fig:netest_tikz}. Inputs are given in blue \textcolor{cyan!20}{\rule{10pt}{10pt}}, (intermediate) outputs in yellow \textcolor{yellow!20}{\rule{10pt}{10pt}}, and the neural network layers in green \textcolor{green!20}{\rule{10pt}{10pt}}. NetEst uses an adversarial training approach to balance the representations $\mathbf{\phi}_i$ \citep{berrevoets2020}. This is achieved by using two discriminators $d_T$ and $d_Z$ that predict the treatment $t_i$ and the peer exposure $z_i$, respectively. 
To balance the representations, the distributions $p(T_i|\mathbf{x}_i,\mathbf{x}_{\mathcal{N}_i})$ and $p(Z_i|\mathbf{x}_i,\mathbf{x}_{\mathcal{N}_i},t_i)$ are uniformed. More specifically, there should be no information in the hidden representation $\phi_i$ that allows to predict $t_i$ or $z_i$. To calculate $\phi_i$, a Graph Convolutional Network (GCN) \citep{kipf2016semi} and an MLP $e_\phi$ are used. $\phi_i$ is fed into the predictor $p_Y$ to make a prediction for the potential outcome ${y}_i$.
The discriminators $d_Z$ and $d_T$, and the predictor $p_Y$ are MLPs with ReLU non-linearities that consist of 3 layers. $e_{\phi}$ has one layer with a ReLU non-linearity. The GCN consists of one convolutional layer \citep{jiang2022estimating}. 

Every epoch consists of two steps. First, the discriminators are trained to accurately predict $T_i$ and $Z_i$. To this end, Binary Cross Entropy (BCE) and Mean Squared Error (MSE) are used to respectively update the discriminators $d_T$ and $d_Z$. In the second step, the discriminators are frozen. Then, regularization losses are calculated based on the discriminators as follows:
\begin{equation}
\mathcal{L}_{uT} = \frac{1}{N}\sum_{i \in \mathcal{N}}(\hat{t}_i - 0.5)^2 \text{ and } \mathcal{L}_{uZ} = \frac{1}{N}\sum_{i \in \mathcal{N}}(\hat{z}_i - c_i)^2.
\end{equation}
These regularization losses are employed to uniform the distributions. Since $T_i$ is binary, we can use $0.5$ as the uniform distribution. For $Z_i$, $c_i$ is sampled uniformly at random from $\left[0,1\right]$ for each $i$ in each iteration. The outcome loss $\mathcal{L}_Y$ compares the prediction $\hat{y}_i$ with the actual outcomes $y_i$ using BCE. $\mathcal{L}_Y$ is used to update the estimator $e_Y$. Finally, the combined loss is calculated: \[\mathcal{L}_{comb} = \mathcal{L}_Y + \alpha \mathcal{L}_{uT} + \gamma \mathcal{L}_{uZ}.\]
This loss is used to update the GCN and the encoder $e_{\phi}$ \citep{jiang2022estimating}. 


The main modifications made to the original NetEst architecture are the change to a binary outcome $Y_i$ and the addition of the encoder $e_\phi$. The latter improves performance by separating $\mathbf{x}_i$ from $\mathbf{x}_{\mathcal{N}_i}$.
Following \citet{jiang2022estimating} we set $\alpha$ and $\gamma$ to 0.5. The Adam optimizer \citep{KingBa15} is employed to train the network. Furthermore, the neural network is trained using full-batch training. Code provided by \citet{jiang2022estimating} was used as a starting point for both the DGP and NetEst implementation.

\subsubsection{Optimization algorithms} To maximize TTE we use two heuristics: the \textit{greedy algorithm (OTAPI-GR)} and the \textit{genetic algorithm (OTAPI-GA)}.

The greedy algorithm is an iterative approach where the allocation set is sequentially expanded 
until the budget $k$ is reached. In each iteration, the marginal gain in TTE associated with the addition of each remaining node to the allocation set is calculated. This marginal gain can be estimated as $\sum_{i \in \mathcal{N}} \hat{\omega}_i(t_i',z_i') -\sum_{i \in \mathcal{N}} \hat{\omega}_i(t_i,z_i) $
where $\mathbf{t}$ is the treatment vector from the previous iteration, and $\mathbf{t'}$ is $\mathbf{t}$ with an additional node $j$ receiving treatment. The treatment vector $\mathbf{t'}$ with the highest estimated TTE is chosen for that iteration.

The Genetic Algorithm (GA) iteratively improves a population of solutions, represented by treatment allocation vectors $\mathbf{t}$, by applying genetic operators such as crossover and mutation. \textcolor{red}{The fitness function used by the GA is the TTE obtained by using the trained causal model.}
To incorporate the budget constraint, the fitness function returns zero when this constraint is violated. An initial set of solutions has to be generated. To converge more quickly, and to get better solutions, the degree and single discount heuristic solutions are added to the initial population. The other solutions in the initial population are randomly generated. Thereafter, the best solutions are selected to be parents to create new solutions (children). These solutions differ from their parents by using crossover (a child is made up of the parts of its parents) and mutation (randomly changing small parts of the child). Thanks to this randomness, escaping from local optima is made possible. Finally, to avoid losing any particularly good solutions, the best solutions from the previous iteration are kept unchanged, and the next iteration begins \citep{gad2023pygad,bucur2016influence}. The GA was implemented using the PyGAD library
\citep{gad2023pygad}. The hyperparameters were tuned using the predicted TTE on the synthetic data with $\beta_{spillover}=0.3$. The hyperparameters are shown in \Cref{tab:GA_hyper}. The number of generations is dependent on the budget $k$ because the algorithm converges faster for lower $k$.

\begin{table}[h]
    \centering
    \begin{tabular}{ll} 
        Parameter& Value\\
        \midrule
        Num. generations& $\begin{cases}
       37 \cdot k + 300 & \text{if } k \leq 100\\
        5,000 & \text{if } k > 100
    \end{cases}$\\
        Population size & $40$\\
        Num. elites  & $5$\\
        Num. parents mating  & $15$\\
        Num. genes mutated  & $1$\\
        Crossover type  & uniform\\
        \bottomrule 
    \end{tabular}
    \caption{Hyperparameters GA.}
    \label{tab:GA_hyper}
\end{table}


\subsubsection{Methods for comparison} 

We implemented the following methods to compare OTAPI with:
\begin{itemize}[leftmargin=*, nosep]
\item \textit{Degree (DEG)}: pick the $k$ nodes with highest degree.
\item \textit{Single Discount (SD)}: stepwise degree as described by \citet{chen2009efficient}. In each iteration, the node with the highest degree is added to the allocation set. Once a node is selected, all its edges are deleted and the degree is recalculated. This process stops when the budget $k$ is reached.
    
\item \textit{CELF}: IM approach. A variation of the greedy algorithm that reduces the required amount of MC simulations per iteration by leveraging the submodularity property that some diffusion processes exhibit \citep{leskovec2007cost}. We employ the Independent Cascade (IC) model with probability $p=0.01$ as the diffusion process because of its popularity within the IM literature \citep{kempe2003maximizing}. We simulate the diffusion process $1000$ times per step to get an accurate estimate of the TTE. 
\item \textit{TARNet}: UM approach. TARNet \citep{shalit2017estimating}, a commonly used causal estimator, is used for ITE estimation. TARNet is trained on individual data only, not leveraging any network information. The $k$ entities with the highest ITE estimates are allocated treatment. We use the implementation for TARNet from \citet{jiang2022estimating} and largely follow their setup. The model is trained using full-batch training with BCE loss and employs the Adam optimizer \cite{KingBa15}. 
\item \textcolor{red}{\textit{Oracle Greedy (OG)}}: To get an idea of the \textcolor{red}{best possible performance} on the evaluation metrics, we implement a method that leverages the true DGP, which simulates the performance of a perfect model $\mathcal{M}$. Since the allocation problem is---even with a perfect model---still \textcolor{red}{an NP-hard} problem, a heuristic \textcolor{red}{is} used to solve it. To this end, we use the greedy heuristic. \textcolor{red}{While there is no guarantee that this solution is close to the true optimal treatment allocation, it serves as a useful baseline to provide insight into the performance of the different methods.}
\end{itemize}

\subsubsection{Evaluation metrics} We define Liftup in TTE for method $m$ and budget $k$ as:
\begin{equation}
\text{Liftup}_m(k) = \frac{\text{TTE}_{m}(k)}{\text{TTE}_{random}(k)}.
\end{equation}
This metric captures the relative increase in TTE of a method compared to randomly assigning an equal amount of treatments. Since there are many possible random assignments, we randomly sample 100 possible assignments for a budget $k$ and calculate the average TTE to estimate $\text{TTE}_{random}(k)$. 100 was chosen because it gave stable estimates for all $k$. 
Moreover, to get a more complete understanding of the results, the relative increase in sum of expected outcomes (RISEO) is defined as:
\begin{equation}
\text{RISEO}_m(k) = \frac{\sum_{i\in \mathcal{N}} \mathbb{E}\bigl[Y_i(t_{i,m},z_{i,m})\bigr]}{\sum_{i\in \mathcal{N}} \mathbb{E}\bigl[Y_i(t_{i,random},z_{i,random})\bigr]}.
\end{equation}
Here, $t_{i,m}$ and $z_{i,m}$ result from the treatment allocation by method $m$. This metric quantifies the relative increase in the sum of expected outcomes (e.g., total sales) compared to random treatment allocation. We use the same strategy to estimate the value of random treatment allocation as in Liftup. 

\subsubsection{Hyperparameter selection}
We use grid search to find the best hyperparameters for TARNet and NetEst for each setup. To select the best model, we use the validation BCE to compare the factual outcomes $y_i$ to the predicted outcomes $\hat{y}_i$. However, the model that is best at predicting the potential outcomes, is not necessarily the best at predicting causal effects. Although model selection in causal inference is still an unresolved issue, better approaches may exist \citep{curth2023search}. However, given that these 
approaches have not yet been tested in a network context, we leave this for future research. The hyperparameter ranges for NetEst and TARNet are shown in \Cref{tab:Netest_hyper,tab:CFR_hyper}. \textcolor{red}{Given the high feature dimensionality in the Enron dataset, we also select the best hidden layer dimension using a grid search over $\{8,16,32\}$ for both NetEst and TARNet. For all other datasets, the hidden layer dimension is set to 16.} The source code for all our experiments can be found in this repository:  \href{https://github.com/daan-caljon/OTAPI}{https://github.com/daan-caljon/OTAPI}.

\begin{table}[h]
    \centering
    \begin{tabular}{ll} 
        Parameter& Range\\
        \midrule
        Learning rate & $\{5\cdot10^{-3},1\cdot10^{-3},5\cdot10^{-4}\}$\\
        Epochs & $\{200,500,800\}$\\
        \bottomrule 
    \end{tabular}
    \caption{Hyperparameters NetEst.}
    \label{tab:Netest_hyper}
\end{table}

\begin{table}[h]
    \centering
    \begin{tabular}{ll} 
        Parameter& Range\\
        \midrule
        Learning rate & $\{5\cdot10^{-3},1\cdot10^{-3},5\cdot10^{-4}\}$\\
        Epochs & $\{500,800\}$\\
        Num. representation layers  & $\{1,2\}$\\
         Num. hypothesis layers  & $\{1,2\}$\\
        \bottomrule 
    \end{tabular}
    \caption{Hyperparameters TARNet.}
    \label{tab:CFR_hyper}
\end{table}

\subsection{Results and discussion}
\begin{figure}[ht]
 \centering
 \subfigure[Flickr]{\includegraphics[width=0.3\textwidth]{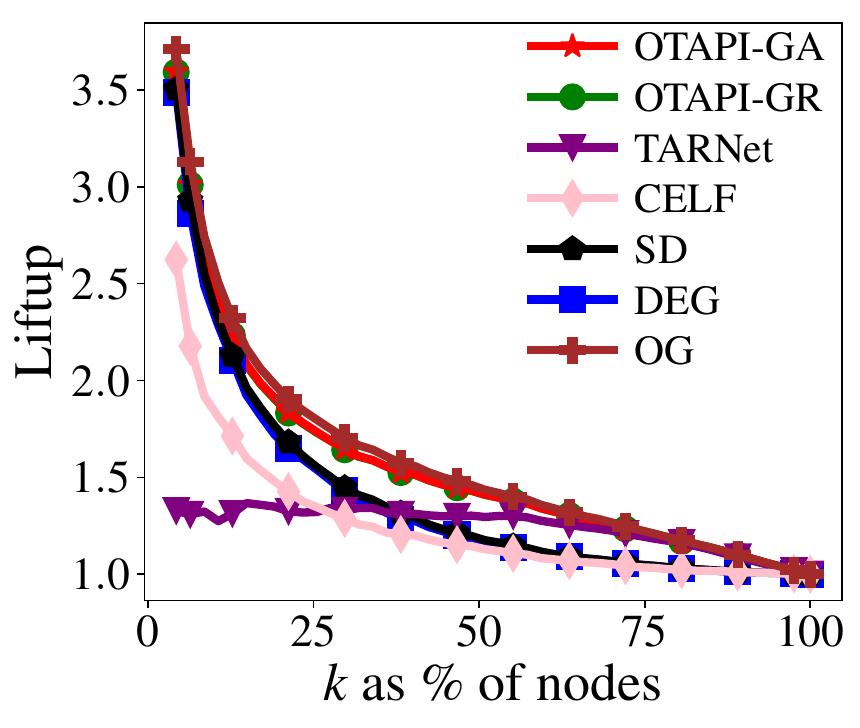}
 \label{fig:Flickr Liftup}}
 \qquad
 \subfigure[BC]{\includegraphics[width=0.3\textwidth]{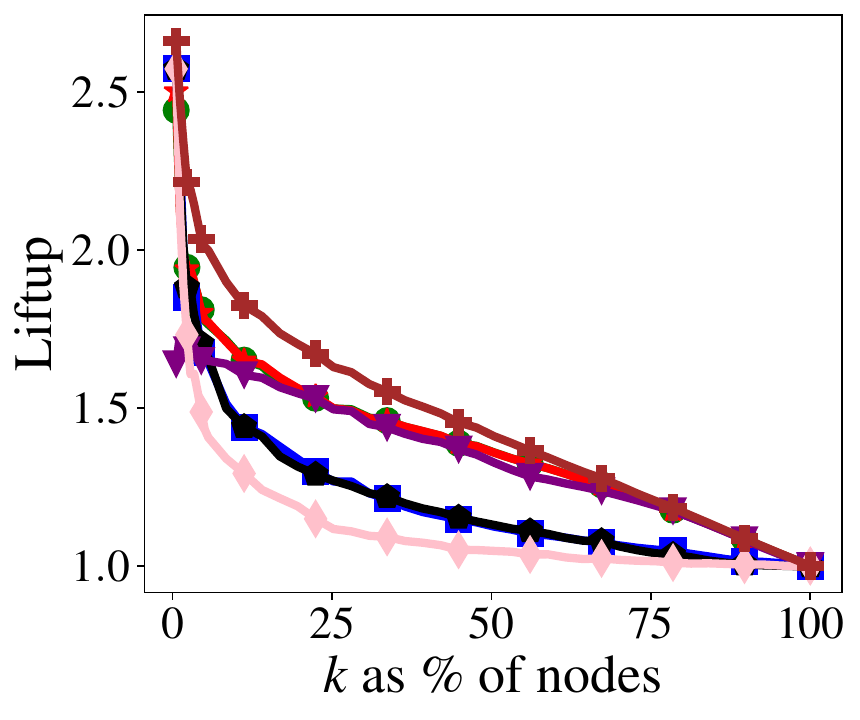}
 \label{fig:BC Liftup}}
\qquad
 \centering
 \subfigure[Flickr]{\includegraphics[width=0.3\textwidth]{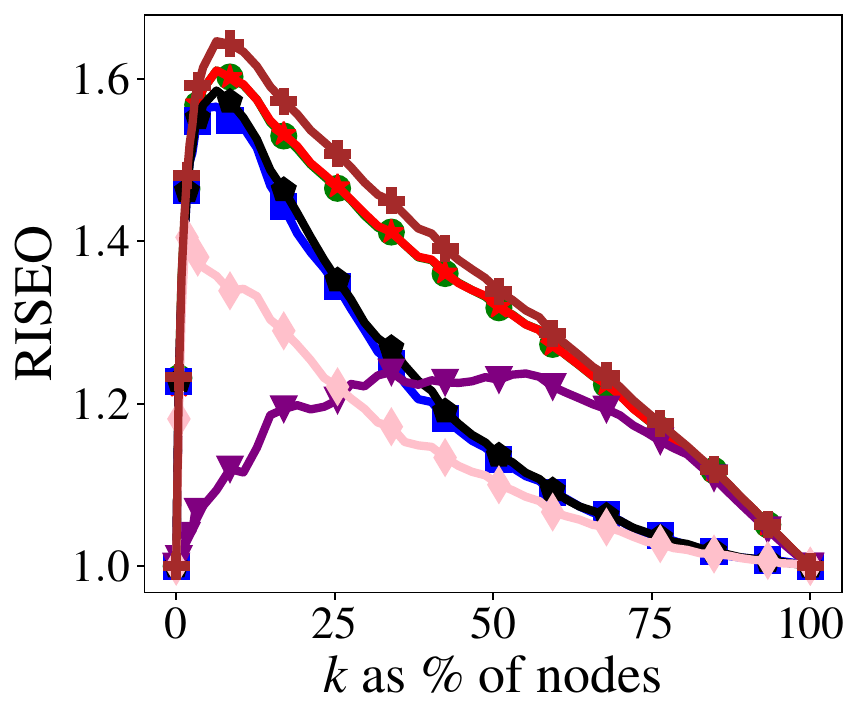}
 \label{fig:Flickr RISEO}}
 \qquad
 \subfigure[BC]{\includegraphics[width=0.3\textwidth]{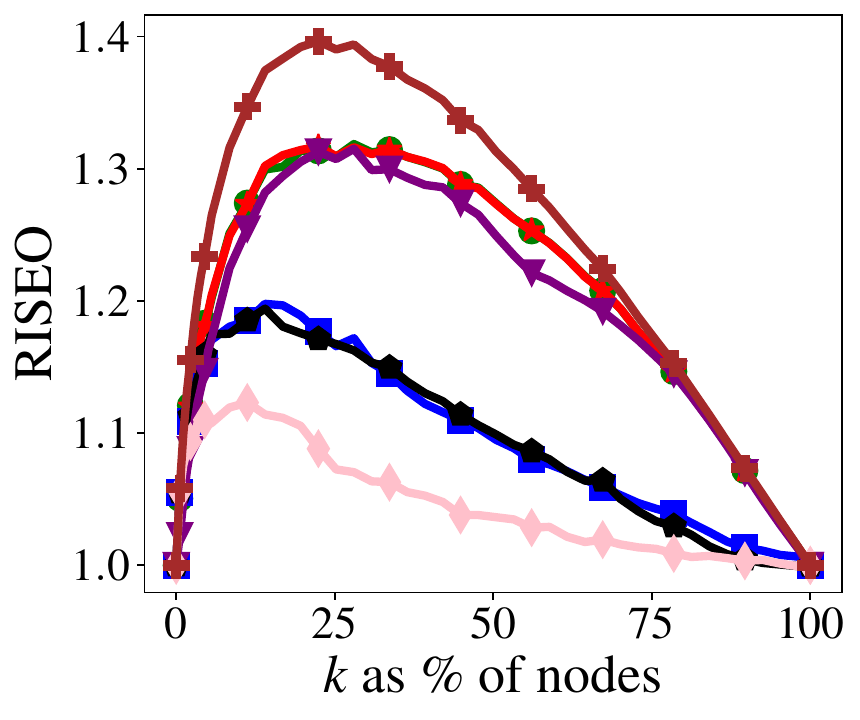}
 \label{fig:BC RISEO}}
 \qquad
 \caption{{Plots for Liftup against $k$ (\subref{fig:Flickr Liftup} and \subref{fig:BC Liftup}) and RISEO against $k$ (\subref{fig:Flickr RISEO} and \subref{fig:BC RISEO}). \textcolor{red}{Higher values are better for both Liftup and RISEO.} OTAPI-GA and OTAPI-GR (less visible because it performs almost identically to OTAPI-GA) outperform the baselines for all $k$ and are very close to \textcolor{red}{OG}. Moreover, TARNet performs very poorly for low $k$, but improves compared to the other baseline methods as $k$ increases. Note that for relatively small $k$ the Liftup may be very high. Because of this, part of the Liftup curve of Flickr has been cut off for better interpretability of the figures.}}
 \label{fig:semi-synthetic}
\end{figure}
\subsubsection{Influence of budget $k$} In \Cref{fig:semi-synthetic}, we show the results for the test sets of the two semi-synthetic datasets BC and Flickr. We plot {Liftup and RISEO against the budget $k$} and observe that \textit{OTAPI-GR and OTAPI-GA outperform (or have at least the same performance as) all other methods for all $k$}. Furthermore, their performance is close to \textcolor{red}{the OG solution}, especially for the Flickr dataset. Note that OTAPI-GR and OTAPI-GA have an almost identical performance (making it hard to see the curves of OTAPI-GR in the figures). Moreover, ``network-aware" methods (i.e., all methods except for TARNet) outperform the UM approach for relatively small budgets $k$. However, once a certain budget is surpassed, the TARNet method outperforms CELF, DEG, and SD for both datasets. This happens already for a small budget $k$ for the BC dataset. This pattern is expected since the degree distribution follows a power law \citep{barabasi1999emergence}. This means that a relatively low number of nodes have a high degree. Intuitively, we would expect that treating these nodes should have a large impact on the network as a whole, which is confirmed by these results. This is further explored in \Cref{sec:watts-strogatz}. The UM approach (TARNet) fails to identify these high-impact entities. As $k$ increases, identifying nodes with a large MITE becomes more important than the spillover effects. OTAPI can capture both the spillover effects and MITEs, which explains why it performs well for all $k$. 
Note that the peaks of the Liftup and RISEO curves have no direct implications in terms of decision-making. The optimal $k$ 
depends on the relative costs and benefits of the treatments and their effects, respectively \citep{verbeke2020foundations}.

\begin{figure}[ht]
 \centering
 \subfigure[$k=1\%$]{\includegraphics[width=0.29\textwidth]{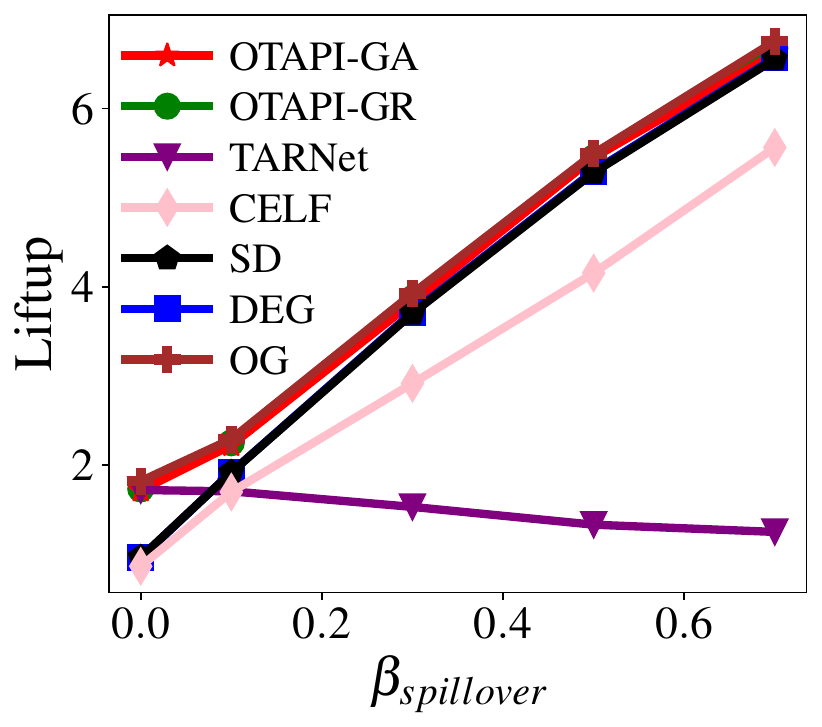}
 \label{fig:1 percent}}
 \quad
 \subfigure[$k=5\%$]{\includegraphics[width=0.3\textwidth]{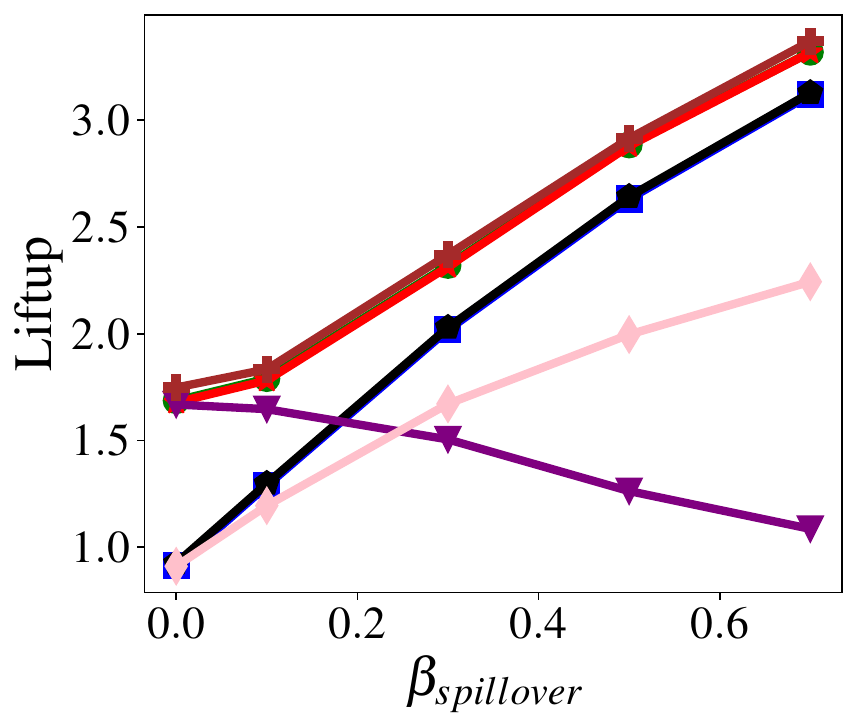}}
 \label{fig:5 percent}
  \quad
 \subfigure[$k=20\%$]{\includegraphics[width=0.3\textwidth]{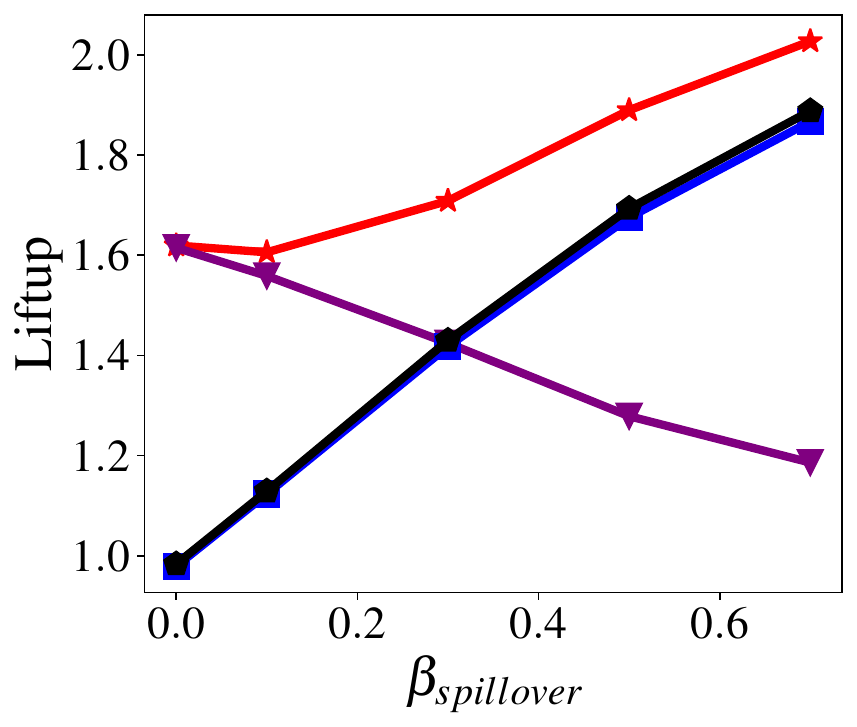}}
 \label{fig:20 percent}
  \quad
 \subfigure[$k=40\%$]{\includegraphics[width=0.3\textwidth]{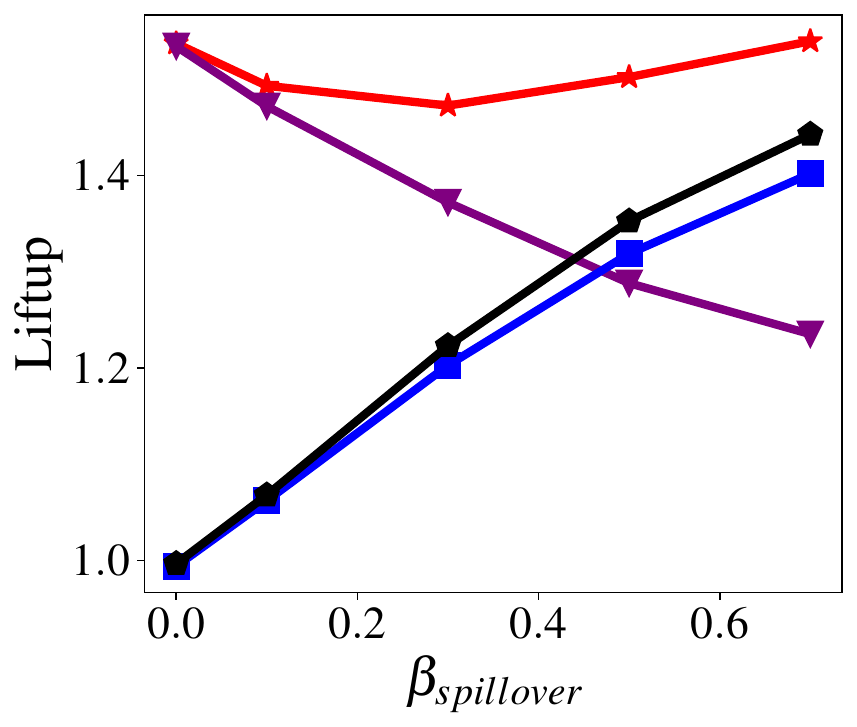}}
 \label{fig:40 percent}
   \quad
 \subfigure[$k=60\%$]{\includegraphics[width=0.3\textwidth]{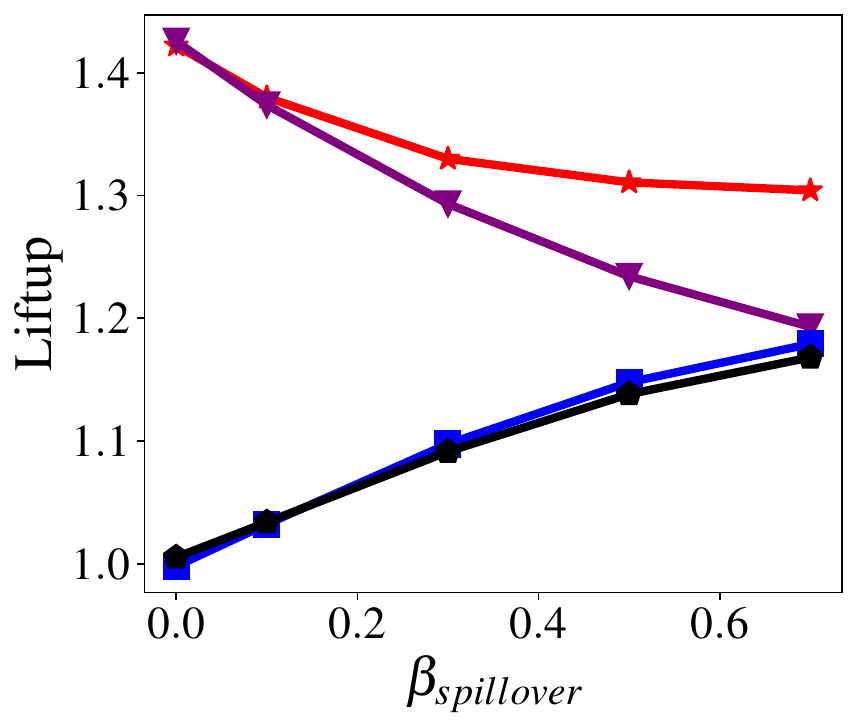}}
 \label{fig:60 percent}
   \quad
 \subfigure[$k=80\%$]{\includegraphics[width=0.3\textwidth]{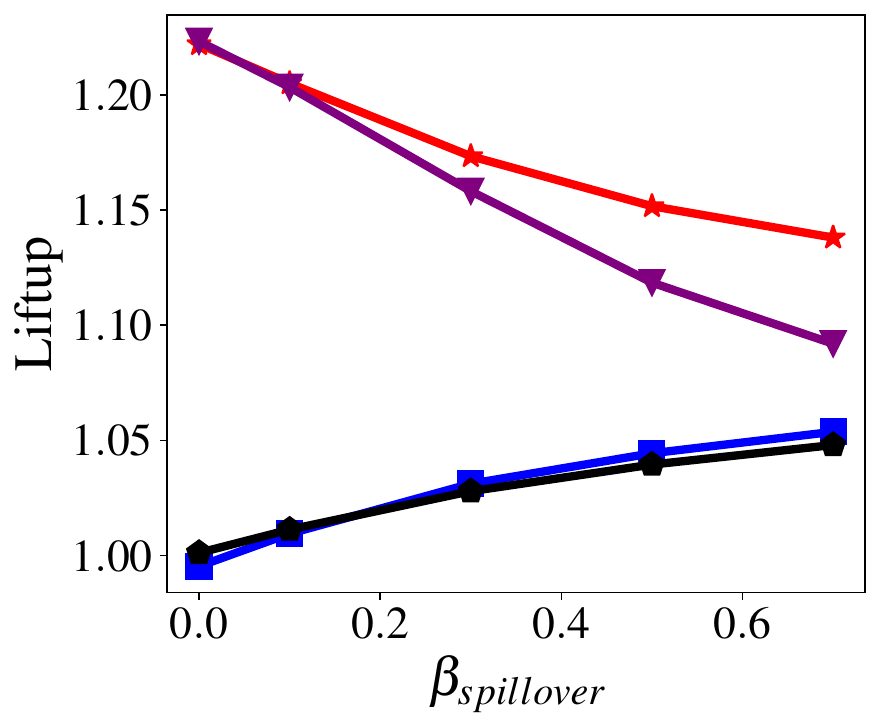}}
 \label{fig:80 percent}
 \caption{{We plot Liftup against $\beta_{spillover}$, which controls the magnitude of the spillover effect. \textcolor{red}{Higher values are better.} OTAPI (GR and GA) outperforms the baseline methods for all values of $\beta_{spillover}$. As $\beta_{spillover}$ increases, the relative performance of the TARNet method becomes worse.}}
 \label{fig:spillover_effect}
\end{figure}

\subsubsection{Influence of the magnitude of the spillover effect}\; An important parameter in the DGP is $\beta_{spillover}$. This parameter determines the average magnitude of the spillover effect. To examine the impact of this parameter on the relative performance of the different methods, we generate synthetic data consisting of 5,000 nodes and 
gradually increase 
$\beta_{spillover}$.
We show the Liftup at different values of $k$ on the test set for all methods for $k=1\%$ and $k=5\%$ and for all non-greedy methods for higher budgets $k$ in \Cref{fig:spillover_effect}. The greedy methods were not run for higher budgets due to a high runtime.
From the results of this experiment, we can draw four interesting conclusions. (1) \textit{OTAPI-GR and OTAPI-GA outperform all benchmarks for all values of 
$\beta_{spillover}$}. (2) The TARNet method performs worse as the spillover effect increases and has the same performance as OTAPI for $\beta_{spillover} = 0$. The latter is expected since this is the classic UM setting with no interference. (3) The relatively simple DEG and SD heuristics exhibit performance more closely aligned with OTAPI as the average spillover effect increases. (4) As already concluded from \Cref{fig:semi-synthetic}, we observe that TARNet becomes more competitive as $k$ increases. Furthermore, DEG and SD perform worse as $k$ increases and do not even outperform TARNet when $k=60\%$ and $k=80\%$ for a large $\beta_{spillover}$.
The fact that OTAPI-GR becomes challenging in terms of runtime as $k$ increases also highlights the advantage of OTAPI-GA compared to OTAPI-GR. Given a certain budget $k$, OTAPI-GA only needs to calculate the best solution for that value of $k$. OTAPI-GR, on the other hand, has to calculate all solutions for smaller $k$ first. 

\begin{figure}[h]
    \centering
    \subfigure[Simulated data]{
        \centering
        \includegraphics[width=0.31\textwidth]{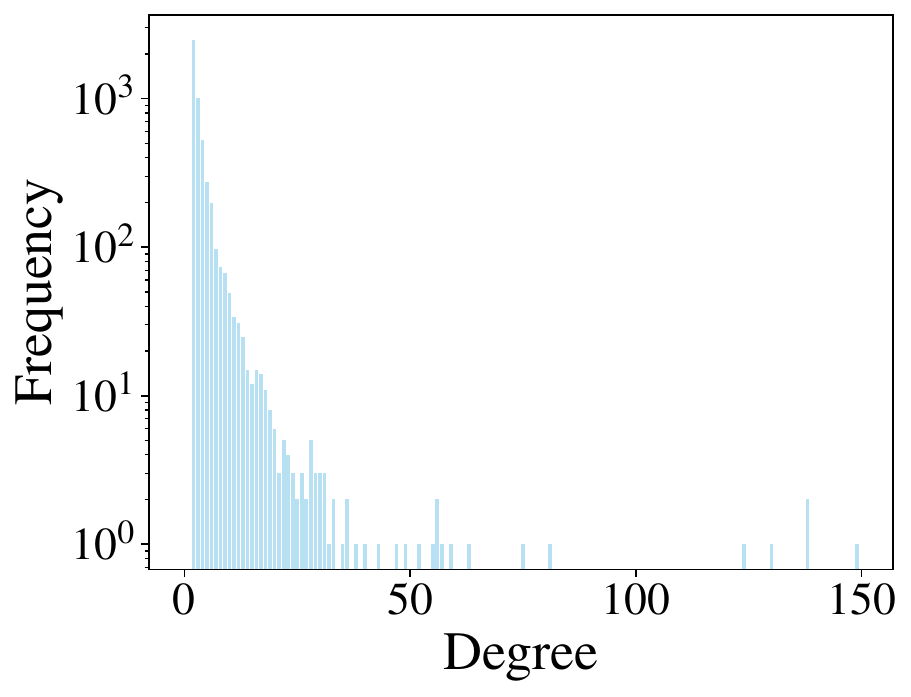}
        \label{fig:Degree simulated}}
    \subfigure[Flickr]{
        \centering
        \includegraphics[width=0.31\textwidth]{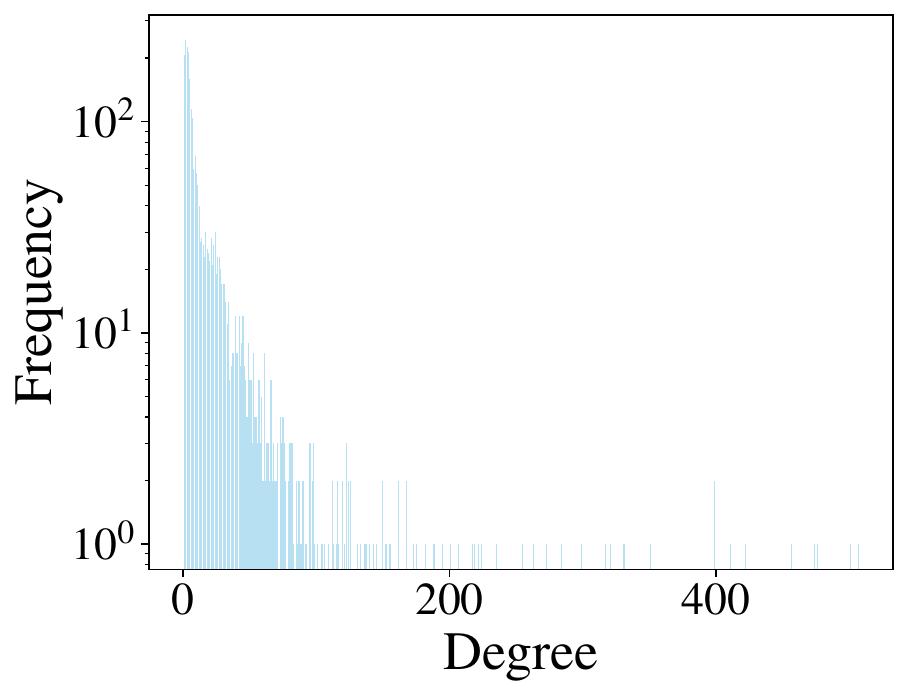}
        \label{fig:degree flickr}}
    \subfigure[BC]{
        \centering
        \includegraphics[width=0.31\textwidth]{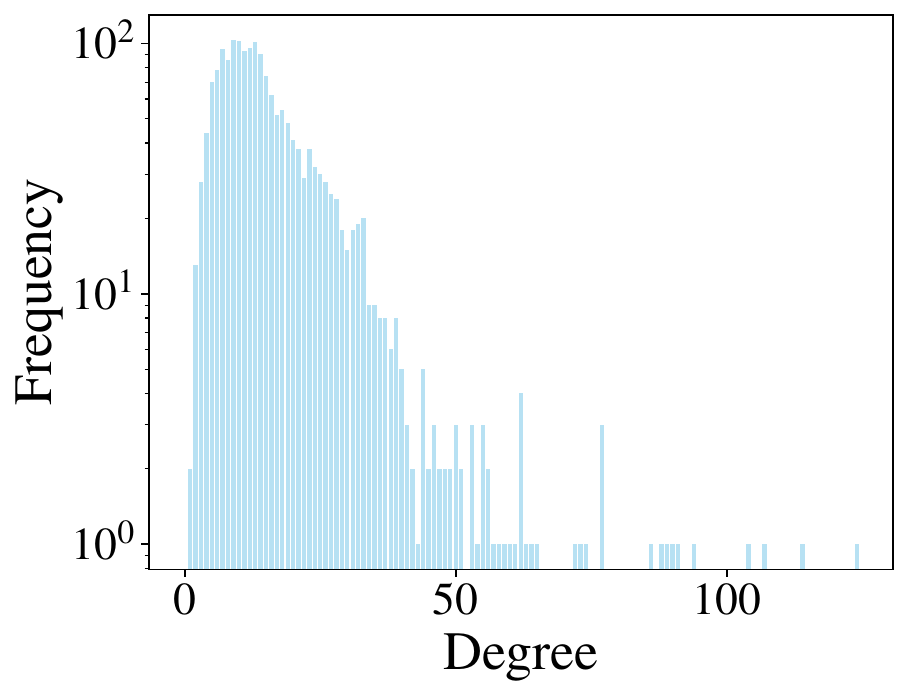}
        \label{fig:degree BC}}%
    \caption{Degree distribution of the different networks. The degree distributions follow a power law, with a few nodes having a very high degree, while the majority has a relatively low degree.}
    \label{fig:degree effect}
\end{figure}

\subsubsection{Impact of the degree distribution} \label{sec:watts-strogatz}
\begin{figure}[ht]
    \centering
\subfigure[Degree distribution]{
        \centering
        \includegraphics[width=0.31\textwidth]{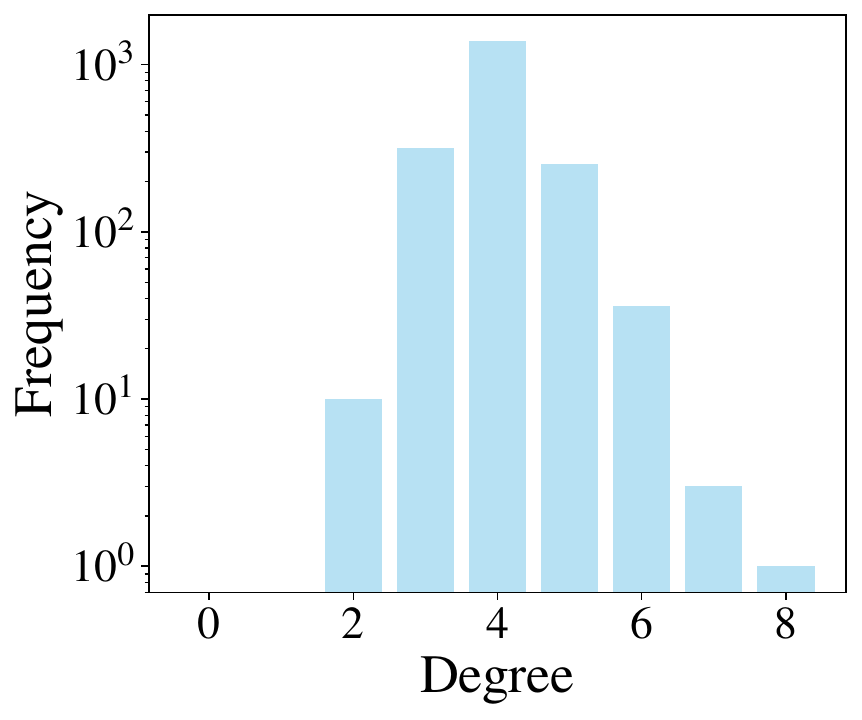}
        \label{fig:Degree WS}}%
    \subfigure[Liftup]{
        \centering
        \includegraphics[width=0.31\textwidth]{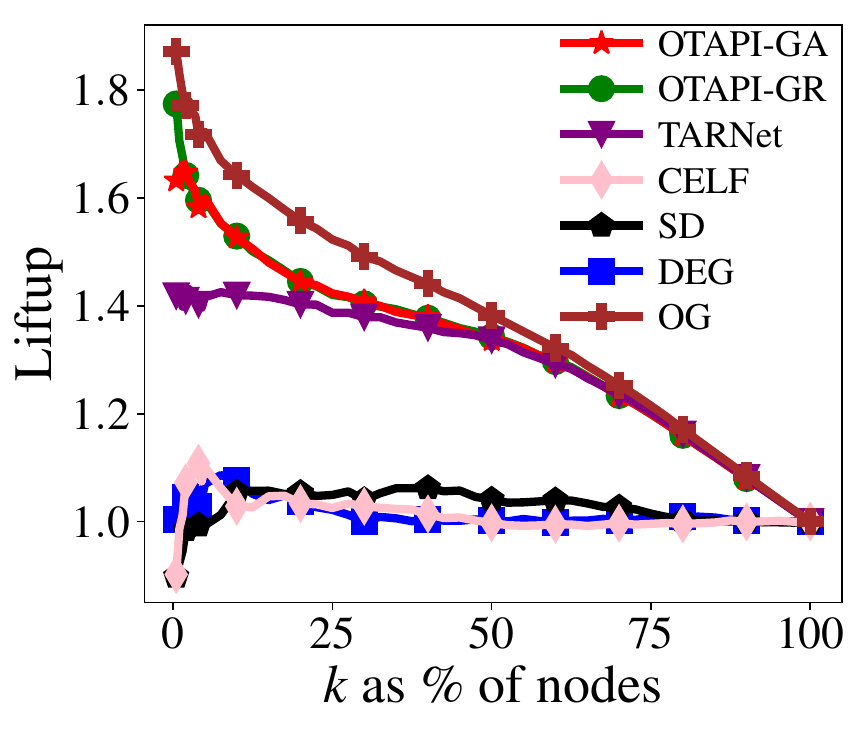}

        \label{fig:liftup WS}}%
    \subfigure[RISEO]{
        \centering
        \includegraphics[width=0.31\textwidth]{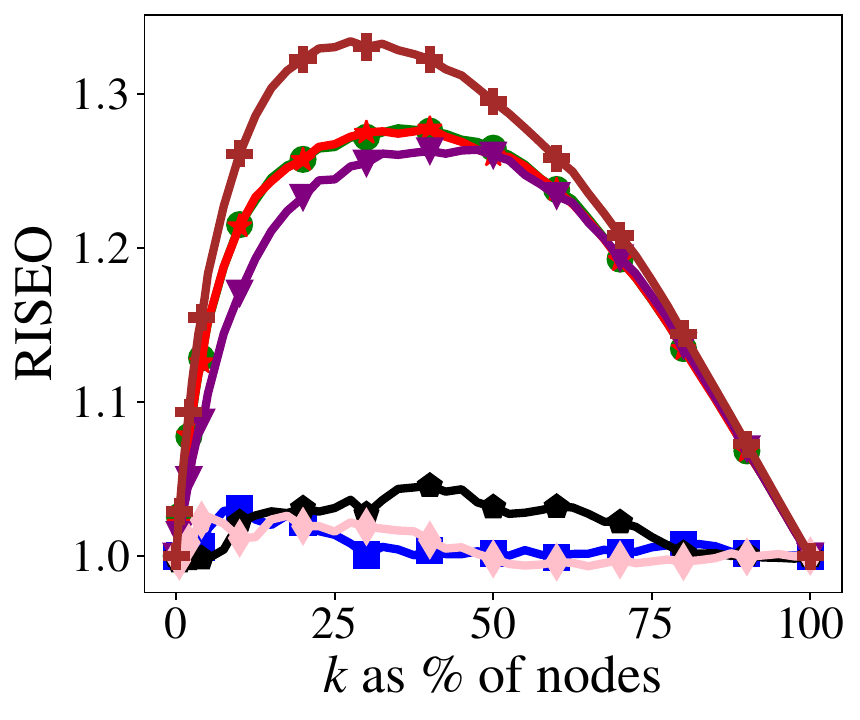}
        \label{fig:influence WS}}%
    \caption{Experiment Watts-Strogatz random network. \textcolor{red}{Higher values are better for both Liftup and RISEO.} In this network, there is no power law for the degree \subref{fig:Degree WS}. Consequently, DEG, SD, and CELF no longer find better treatment allocations than random treatment allocation (i.e., Liftup $\approx 1$). OTAPI still performs well.}
    \label{fig:WS figs}
\end{figure}

All networks investigated in this paper possess the scale-free property \citep{barabasi1999emergence}. This means that the distribution of the degree follows a power law. As shown in \Cref{fig:degree effect}, a small number of nodes has a very high degree. Consequently, they will most likely have a large network effect. This may explain why 
DEG and SD 
perform quite well for small $k$. To further investigate this hypothesis, we simulate a small dataset of 2,000 nodes using the Watts-Strogatz \citep{watts1998collective} random network model. Networks that follow this model do not possess the scale-free property. The degree distribution for the simulated data using the Watt-Strogatz model is shown in Figure \ref{fig:Degree WS}. To make a fair comparison to the Barabási–Albert random networks, the parameters of the Watts-Strogatz random network were chosen to have the same average degree (4).

We set $\beta_{spillover}=0.3$ and run the experiment for all $k$. The results are shown in Figures \ref{fig:liftup WS} and \ref{fig:influence WS}. These differ significantly from previous results. TARNet performs very well even though there is a network effect. It outperforms the methods that only look at network structure (DEG, SD, and CELF). Moreover, these methods are not better than randomly allocating treatments. Although the results for the baseline methods are different, the results for OTAPI remain the same. It still outperforms (or has the same performance) as the best baseline.


\begin{figure}[ht]
    \centering
    \subfigure[$k=5\%$]{
        \centering
        \resizebox{0.31\textwidth}{!}{
            
\newcommand{\colorscale}[1]{%
    \pgfmathsetmacro{\value}{#1}%
    \ifdim \value pt > 0.95 pt
        \cellcolor{green!60}%
    \else
        \ifdim \value pt > 0.90 pt
            \cellcolor{green!55}%
        \else
            \ifdim \value pt > 0.85 pt
                \cellcolor{green!50}%
            \else
                \ifdim \value pt > 0.80 pt
                    \cellcolor{green!45}%
                \else
                    \ifdim \value pt > 0.75 pt
                        \cellcolor{green!40}%
                    \else
                        \ifdim \value pt > 0.70 pt
                            \cellcolor{green!35}%
                        \else
                            \ifdim \value pt > 0.65 pt
                                \cellcolor{green!30}%
                            \else
                                \ifdim \value pt > 0.60 pt
                                    \cellcolor{green!25}%
                                \else
                                    \ifdim \value pt > 0.55 pt
                                        \cellcolor{green!20}%
                                    \else
                                        \ifdim \value pt > 0.50 pt
                                            \cellcolor{green!15}%
                                        \else
                                            \ifdim \value pt > 0.45 pt
                                                \cellcolor{orange!15}%
                                            \else
                                                \ifdim \value pt > 0.40 pt
                                                    \cellcolor{orange!20}%
                                                \else
                                                    \ifdim \value pt > 0.35 pt
                                                        \cellcolor{orange!25}%
                                                    \else
                                                        \ifdim \value pt > 0.30 pt
                                                            \cellcolor{orange!30}%
                                                        \else
                                                            \ifdim \value pt > 0.25 pt
                                                                \cellcolor{orange!35}%
                                                            \else
                                                                \ifdim \value pt > 0.20 pt
                                                                    \cellcolor{orange!40}%
                                                                \else
                                                                    \ifdim \value pt > 0.15 pt
                                                                        \cellcolor{orange!45}%
                                                                    \else
                                                                        \ifdim \value pt > 0.10 pt
                                                                            \cellcolor{orange!50}%
                                                                        \else
                                                                            \ifdim \value pt > 0.05 pt
                                                                                \cellcolor{orange!55}%
                                                                            \else
                                                                                \cellcolor{orange!60}%
                                                                            \fi
                                                                        \fi
                                                                    \fi
                                                                \fi
                                                            \fi
                                                        \fi
                                                    \fi
                                                \fi
                                            \fi
                                        \fi
                                    \fi
                                \fi
                            \fi
                        \fi
                    \fi
                \fi
            \fi
        \fi
    \fi
    #1%
}

\begin{tabular}{lccccccc}  
        \toprule
        \multicolumn{2}{c}{} & \multicolumn{3}{c}{Classic IM} & \multicolumn{1}{c}{UM} & \multicolumn{2}{c}{OTAPI}\\
        \cmidrule(r){3-5} \cmidrule(lr){6-6} \cmidrule(l){7-8}
             & OG                           & DEG                          & SD                           & CELF                         & TAR & GR & GA\\
        \midrule
OG   & \cellcolor[HTML]{70AD47}1.00 &                              &                              &                              &  &  &  \\
DEG  & \cellcolor[HTML]{94C275}0.69 & \cellcolor[HTML]{70AD47}1.00 &                              &                              &  &  &  \\
SD   & \cellcolor[HTML]{8DBE6C}0.75 & \cellcolor[HTML]{7AB353}0.92 & \cellcolor[HTML]{70AD47}1.00 &                              &  &  &  \\
CELF & \cellcolor[HTML]{B9D8A6}0.36 & \cellcolor[HTML]{B8D7A4}0.37 & \cellcolor[HTML]{B8D7A4}0.37 & \cellcolor[HTML]{70AD47}1.00 &  &  &  \\
TAR &
  \cellcolor[HTML]{D9EACF}0.08 &
  \cellcolor[HTML]{DDECD3}0.05 &
  \cellcolor[HTML]{DCECD2}0.06 &
  \cellcolor[HTML]{D8EACD}0.09 &
  \cellcolor[HTML]{70AD47}1.00 &
   &
   \\
GR &
  \cellcolor[HTML]{86BA63}0.81 &
  \cellcolor[HTML]{9EC882}0.60 &
  \cellcolor[HTML]{9BC67E}0.63 &
  \cellcolor[HTML]{BEDAAB}0.32 &
  \cellcolor[HTML]{D7E9CC}0.10 &
  \cellcolor[HTML]{70AD47}1.00 &
   \\
GA &
  \cellcolor[HTML]{85B962}0.82 &
  \cellcolor[HTML]{9DC781}0.61 &
  \cellcolor[HTML]{9AC57C}0.64 &
  \cellcolor[HTML]{BCD9A9}0.34 &
  \cellcolor[HTML]{D7E9CC}0.10 &
  \cellcolor[HTML]{74AF4C}0.97 &
  \cellcolor[HTML]{70AD47}1.00 \\
        \bottomrule
    \vspace{0.2cm}
\end{tabular}
    \vspace{0.2cm}

        }

    }
    \subfigure[$k=20\%$]{
        \centering
        \resizebox{0.31\textwidth}{!}{
            \begin{tabular}{lccccccc}  
        \toprule
        \multicolumn{2}{c}{} & \multicolumn{3}{c}{Classic IM} & \multicolumn{1}{c}{UM} & \multicolumn{2}{c}{OTAPI}\\
        \cmidrule(r){3-5} \cmidrule(lr){6-6} \cmidrule(l){7-8}
             & OG                           & DEG                          & SD                           & CELF                         & TAR & GR & GA\\
        \midrule
OG   & \cellcolor[HTML]{70AD47}1.00 &                              &                              &                              &  &  &  \\
DEG  & \cellcolor[HTML]{A0C985}0.58 & \cellcolor[HTML]{70AD47}1.00 &                              &                              &  &  &  \\
SD   & \cellcolor[HTML]{9EC882}0.60 & \cellcolor[HTML]{82B75E}0.85 & \cellcolor[HTML]{70AD47}1.00 &                              &  &  &  \\
CELF & \cellcolor[HTML]{B7D6A3}0.38 & \cellcolor[HTML]{ABCF92}0.49 & \cellcolor[HTML]{AED197}0.46 & \cellcolor[HTML]{70AD47}1.00 &  &  &  \\
TAR &
  \cellcolor[HTML]{AED197}0.46 &
  \cellcolor[HTML]{CBE2BC}0.21 &
  \cellcolor[HTML]{CCE2BD}0.20 &
  \cellcolor[HTML]{C7E0B7}0.24 &
  \cellcolor[HTML]{70AD47}1.00 &
   &
   \\
GR &
  \cellcolor[HTML]{93C174}0.70 &
  \cellcolor[HTML]{ADD095}0.47 &
  \cellcolor[HTML]{ACD094}0.48 &
  \cellcolor[HTML]{B8D7A4}0.37 &
  \cellcolor[HTML]{A6CD8D}0.53 &
  \cellcolor[HTML]{70AD47}1.00 &
   \\
GA &
  \cellcolor[HTML]{92C172}0.71 &
  \cellcolor[HTML]{ADD095}0.47 &
  \cellcolor[HTML]{ABCF92}0.49 &
  \cellcolor[HTML]{B7D6A3}0.38 &
  \cellcolor[HTML]{A6CD8D}0.53 &
  \cellcolor[HTML]{75B04D}0.96 &
  \cellcolor[HTML]{70AD47}1.00 \\
        \bottomrule
\vspace{0.2cm}
\end{tabular}
        \vspace{0.2cm}

        }  
    }
    \subfigure[$k=40\%$]{
        \centering
        \resizebox{0.31\textwidth}{!}{
            \begin{tabular}{lccccccc}  
        \toprule
        \multicolumn{2}{c}{} & \multicolumn{3}{c}{Classic IM} & \multicolumn{1}{c}{UM} & \multicolumn{2}{c}{OTAPI}\\
        \cmidrule(r){3-5} \cmidrule(lr){6-6} \cmidrule(l){7-8}
             & OG                           & DEG                          & SD                           & CELF                         & TAR & GR & GA\\
        \midrule
OG   & \cellcolor[HTML]{70AD47}1.00 &                              &                              &                              &  &  &  \\
DEG  & \cellcolor[HTML]{A4CB8A}0.55 & \cellcolor[HTML]{70AD47}1.00 &                              &                              &  &  &  \\
SD   & \cellcolor[HTML]{A3CB88}0.56 & \cellcolor[HTML]{85B962}0.82 & \cellcolor[HTML]{70AD47}1.00 &                              &  &  &  \\
CELF & \cellcolor[HTML]{ADD095}0.47 & \cellcolor[HTML]{9FC984}0.59 & \cellcolor[HTML]{ADD095}0.47 & \cellcolor[HTML]{70AD47}1.00 &  &  &  \\
TAR &
  \cellcolor[HTML]{8FBF6F}0.73 &
  \cellcolor[HTML]{B5D5A0}0.40 &
  \cellcolor[HTML]{B5D5A0}0.40 &
  \cellcolor[HTML]{B4D49E}0.41 &
  \cellcolor[HTML]{70AD47}1.00 &
   &
   \\
GR &
  \cellcolor[HTML]{86BA63}0.81 &
  \cellcolor[HTML]{A7CD8E}0.52 &
  \cellcolor[HTML]{A4CB8A}0.55 &
  \cellcolor[HTML]{ADD095}0.47 &
  \cellcolor[HTML]{86BA63}0.81 &
  \cellcolor[HTML]{70AD47}1.00 &
   \\
GA &
  \cellcolor[HTML]{86BA63}0.81 &
  \cellcolor[HTML]{A7CD8E}0.52 &
  \cellcolor[HTML]{A3CB88}0.56 &
  \cellcolor[HTML]{ADD095}0.47 &
  \cellcolor[HTML]{86BA63}0.81 &
  \cellcolor[HTML]{73AF4A}0.98 &
  \cellcolor[HTML]{70AD47}1.00\\
        \bottomrule
    \vspace{0.2cm}
\end{tabular}
    \vspace{0.2cm}
        }  
    }
    \caption{Allocation similarity for the different methods. Allocation similarity is defined as the ratio of the number of selected nodes that two methods have in common to the budget $k$.}
    \label{tab:sim_matrix}
\end{figure}

\subsubsection{Allocation similarity between methods} 
To gain more insight into the similarities and differences between methods, we calculate a similarity matrix between the solutions for the Flickr test set at $k=5\%$, $k=20\%$, and $k=40\%$. Similarity is defined as the number of selected nodes that two methods have in common divided by the budget $k$. The similarity matrices are shown in \Cref{tab:sim_matrix}. As expected, solutions with a larger Liftup (see \Cref{fig:Flickr Liftup}) are more similar to the OG solution. Furthermore, we observe that OTAPI-GR and OTAPI-GA yield an almost identical solution. The high similarity explains why the Liftup for both methods is almost identical. This similarity between the two methods persists for higher $k$. Furthermore, the ability to leverage network structure plays a significant role in finding solutions with high similarity to OG. This pattern changes as $k$ increases, then the similarity between OG and TARNet, and OTAPI and TARNet increases, while the similarity between OG and DEG, and OG and SD decreases. This is because, at some point when increasing $k$, the MITEs gradually become more important than the spillover effects. If this analysis were to be extended to higher $k$, it would become increasingly difficult to extract insightful conclusions from the similarity matrix as a random solution would already have an expected similarity of $k\%$ with each solution. Consequently, $k=40\%$ is the highest budget we investigate using the similarity matrix.
\subsubsection{\textcolor{red}{Potential influence of dataset size and feature dimensionality}}

\begin{figure}[ht]
    \centering
\subfigure[Degree distribution]{
        \centering
        \includegraphics[width=0.31\textwidth]{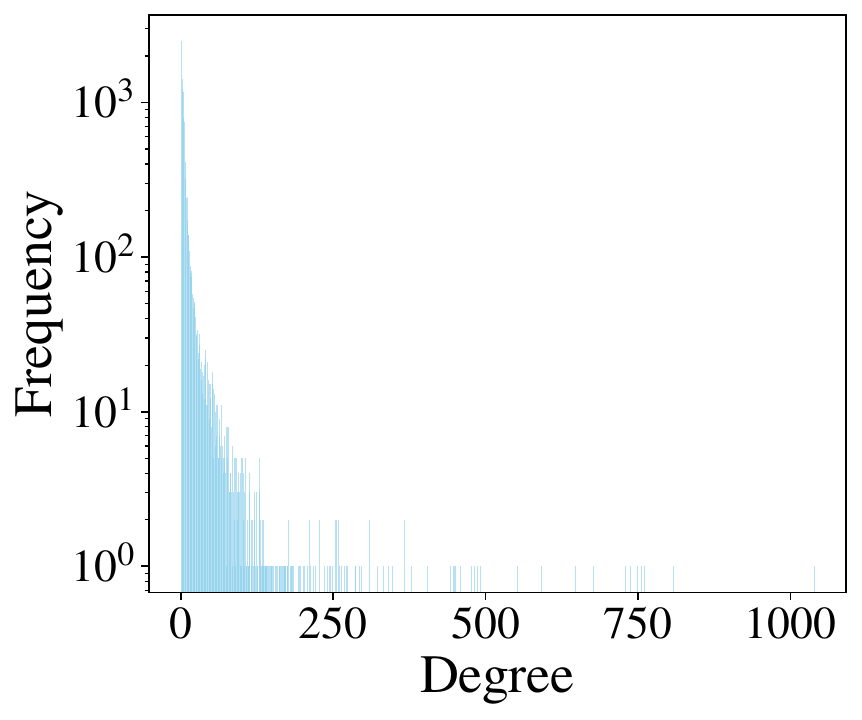}
        \label{fig:Degree Enron}}%
    \subfigure[Liftup]{
        \centering
        \includegraphics[width=0.31\textwidth]{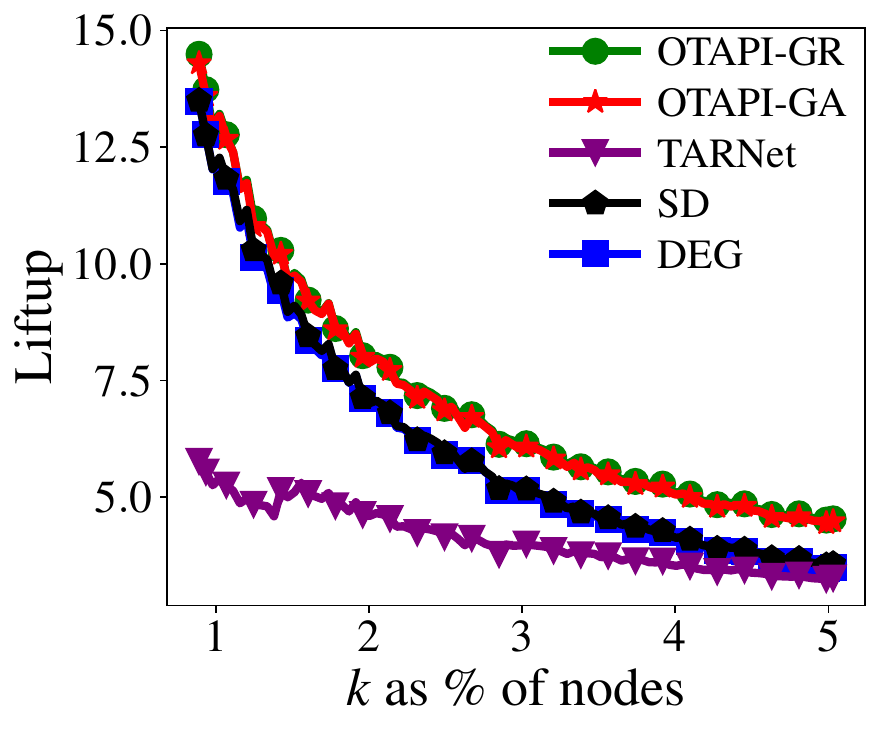}}
        \label{fig:liftup Enron}%
    \subfigure[RISEO]{
        \centering
        \includegraphics[width=0.31\textwidth]{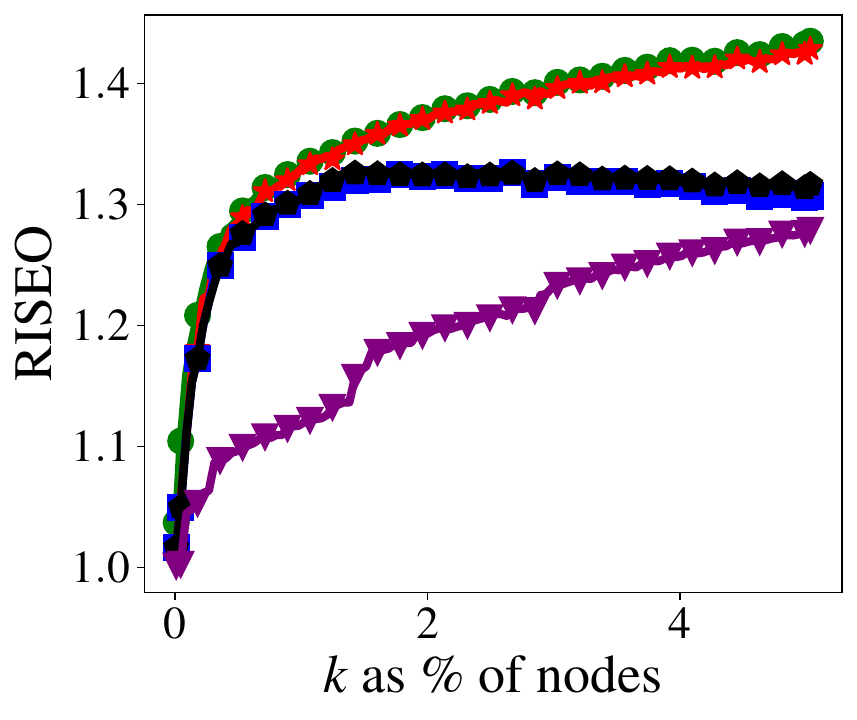}
        \label{fig:influence Enron}}%
    \caption{\textcolor{red}{Liftup and RISEO for the Enron dataset for $k$ up to 5\%.} \textcolor{red}{Higher values are better for both Liftup and RISEO.}}
    \label{fig:Enron figs}
\end{figure}

\textcolor{red}{Given the higher dimensionality and larger size of the Enron dataset, we investigate whether the relative performance of the different methods remains consistent. Due to the large number of nodes, OG and CELF are computationally too expensive and are therefore omitted (see \Cref{sec:runtime}). The results, shown in \Cref{fig:Enron figs}, are similar to those presented for the other datasets. OTAPI achieves the best performance, closely followed by DEG and SD for the lowest budgets. As the budget increases, the performance of DEG and SD deteriorates relative to OTAPI, and at $k=5\%$, their performance is comparable to that of TARNet. This experiment suggests that as both the feature dimensionality and the number of nodes increase, the relative performance of the different methods remains consistent.} 

\subsubsection{\textcolor{red}{Empirical analysis of runtime}}\label{sec:runtime}

\begin{figure}[ht]
    \centering
\subfigure[$k=5\%$]{
        \centering
        \includegraphics[width=0.31\textwidth]{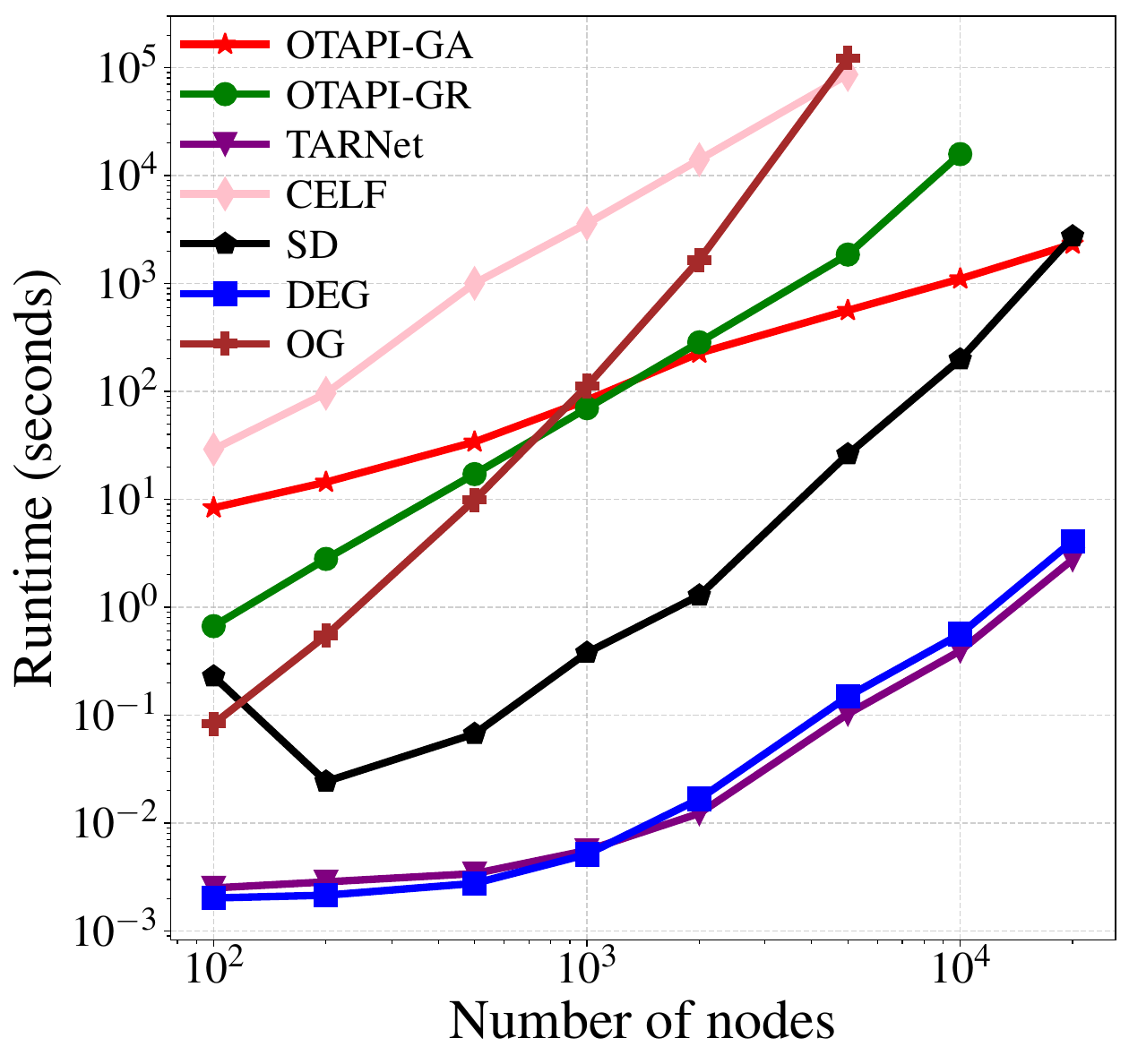}
        \label{fig:runtime 5}}%
    \subfigure[$k=20\%$]{
        \centering
        \includegraphics[width=0.31\textwidth]{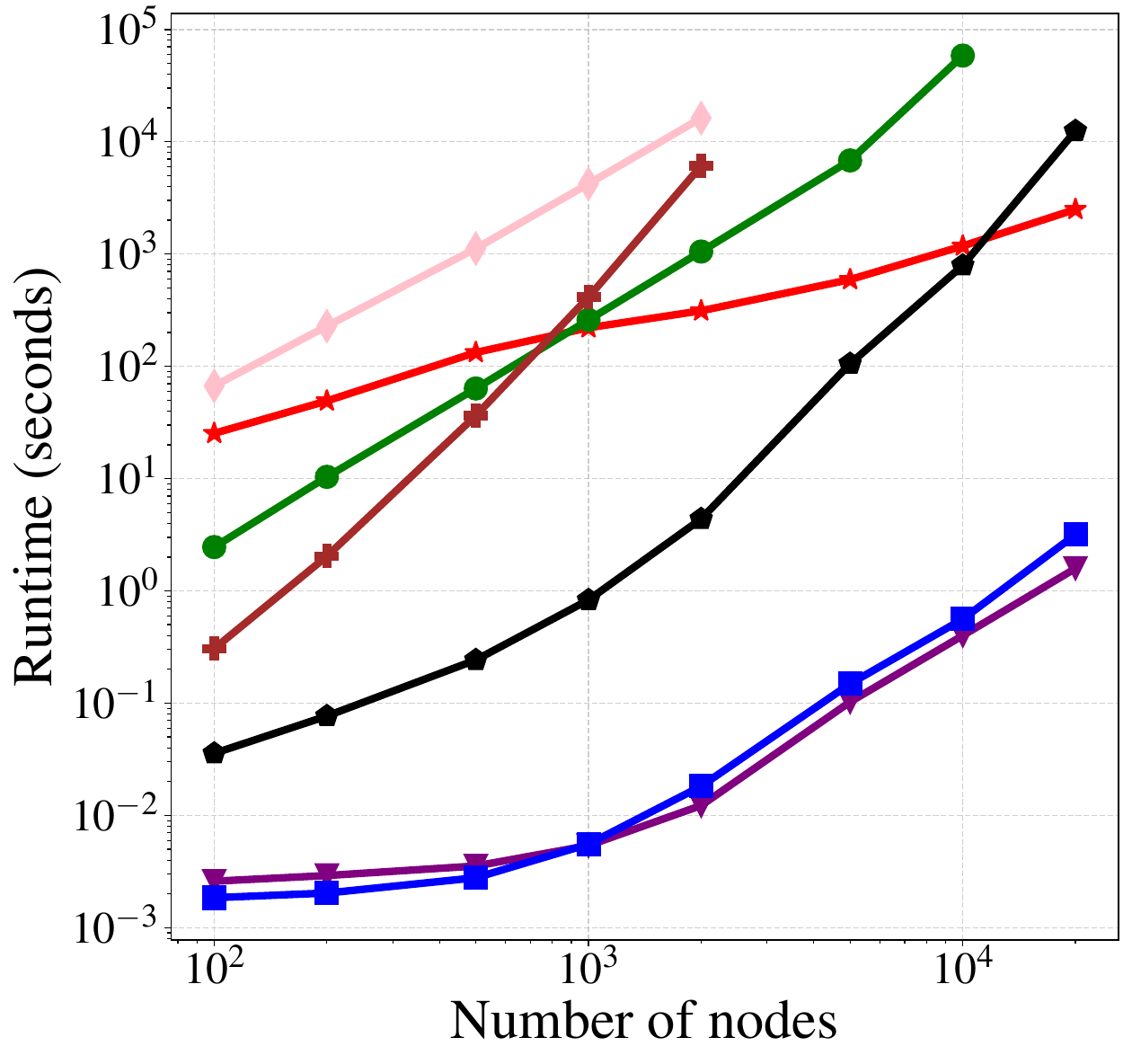}}
        \label{fig:runtime 20}%
    \subfigure[$k=50\%$]{
        \centering
        \includegraphics[width=0.31\textwidth]{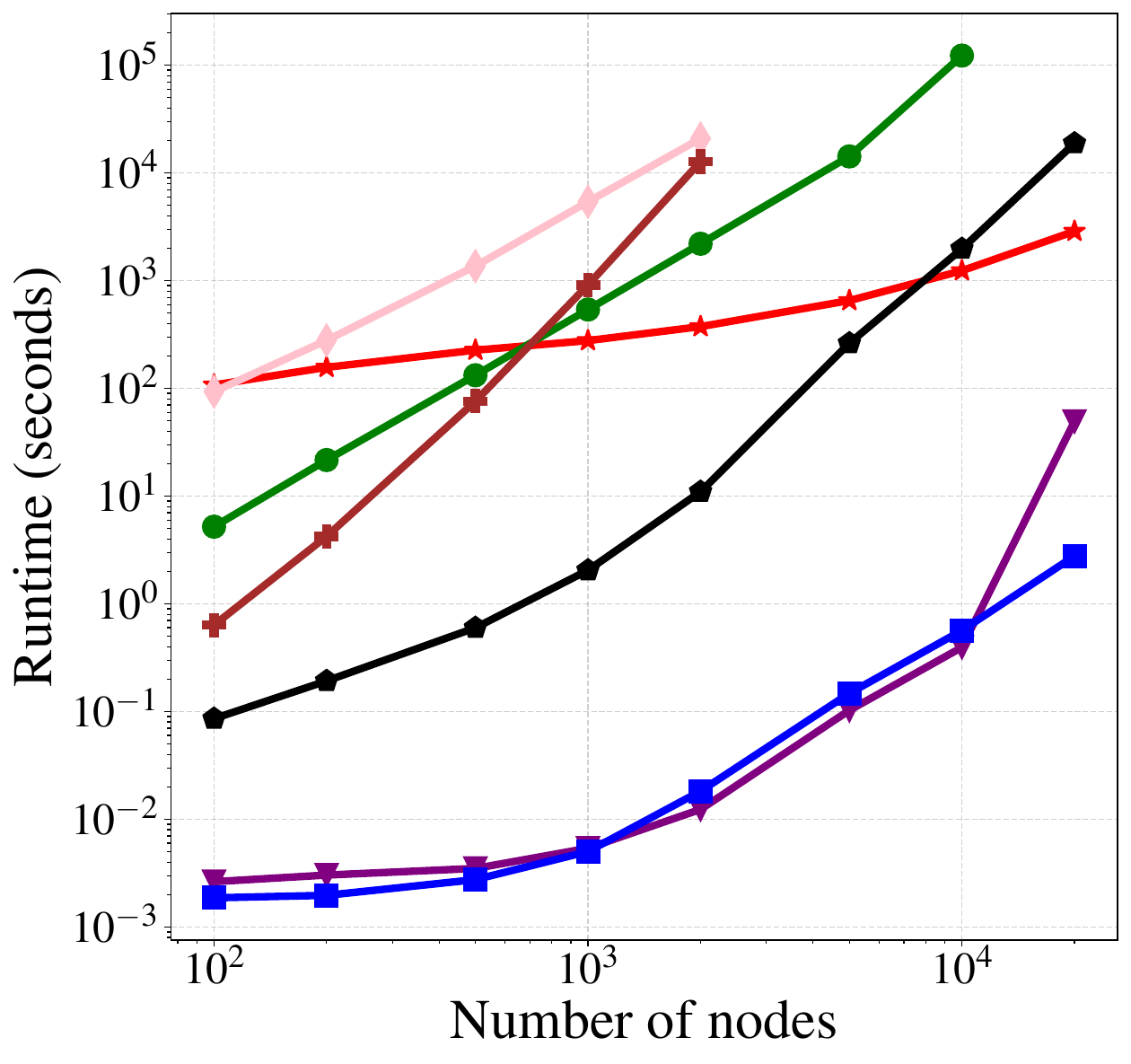}
        \label{fig:runtime 50}}%
    \caption{\textcolor{red}{Empirical runtimes for the different methods on simulated data. Both axes are on a logarithmic scale.}}
    \label{fig:runtime figs}
\end{figure}

\textcolor{red}{In \Cref{fig:runtime figs}, the runtimes of the different methods are shown across a range of simulated dataset sizes for different budgets $k$. Note that both axes in the figures are on a logarithmic scale. All experiments were conducted on a machine with an Intel Xeon Gold 6140 CPU @ 2.3 GHz, 45 GiB of RAM, and an NVIDIA P100 GPU. If an algorithm exceeds the maximum runtime of 48 hours, it is terminated and its runtime is excluded from the figure. The figures show that the methods based on a greedy approach (OTAPI-GR, CELF, and OG) exhibit a substantial increase in runtime as the number of nodes grows. The runtime of the iterative SD approach also increases with dataset size. TARNet and DEG have the lowest runtimes, as they only require calculating a metric (degree or estimated TTE) and then sorting all instances according to this metric. The runtime of OTAPI-GA increases only slightly with dataset size.}

\section{Conclusion}\label{sec: conclusion}
We present OTAPI, a novel method to optimize treatment allocation in the presence of network interference. Currently, two alternative approaches, IM and UM, exist. OTAPI fills the gap between these methods by combining their strengths. Our method consists of two steps. First, a relational causal estimator is trained on observational data to accurately estimate treatment effects in the presence of interference. \textcolor{red}{Second, 
the combinatorial problem of finding the optimal treatment allocation $\mathbf{t}^*$ for a potentially unseen network
is solved 
by using an optimization algorithm that intelligently employs treatment effect estimates from the trained relational causal estimator from the first step.} \textcolor{red}{OTAPI is agnostic to the specific method used in each step; that is, any relational causal estimator and any optimization algorithm from the classic IM literature can be used.}

We have empirically demonstrated that OTAPI outperforms alternative treatment allocation methods. Moreover, the relative performance of our method is robust to both the amount of available treatments $k$ and the magnitude of the spillover effects. By contrast, the relative performance of IM and UM approaches is highly dependent on these parameters. \textcolor{red}{Additional experiments showed that OTAPI maintained its superior relative performance across varying degree distributions, dataset sizes, and feature dimensionalities, outperforming the baselines in every setting considered.}

A limitation of our work is that the assumed causal structure might not fully capture reality, as it does not account for potential contagion effects, which means that the outcome of an entity at time $t$ can influence the outcome of another entity at time $t+1$ \citep{ogburn2014causal}. Moreover, we have followed existing literature \citep{forastiere2021identification,jiang2022estimating} by assuming that the exposure mapping $Z$ is defined as the ratio of neighbors treated. This assumption will likely be violated in practice. Consequently, the causal estimator that was used in this paper may become quite inaccurate. However, since OTAPI is agnostic to the relational causal estimator that is used, other estimators that do not need this assumption \citep[e.g.,][]{huang2023modeling,adhikari2023inferring} can be seamlessly incorporated. Nevertheless, OTAPI's performance using more
complex relational causal estimators should be validated in future work.
Furthermore, the estimation errors from the relational causal estimator might have an impact on the treatment allocation optimization in the second step of OTAPI. Therefore, an interesting avenue for future research is to investigate whether an end-to-end approach, where the treatment allocation is directly learned from the input data, is feasible for this problem. 
Finally, also incorporating potential costs and benefits of treatments and outcomes provides an interesting direction for future research to determine the optimal number of treatments $k^*$ in a data-driven manner \citep{verbeke2020foundations,verbeke2023or}.

\section{Acknowledgements} 
Wouter Verbeke and Daan Caljon thankfully acknowledge the financial support from Research Foundation -- Flanders (project G038723N). Jente Van Belle gratefully acknowledges the financial support from Research Foundation -- Flanders (project 12AZX24N).
\bibliographystyle{elsarticle-harv} 
\bibliography{bibliography}

\newpage
\appendix

\section{\textcolor{red}{Identifiability of causal effects}}\label{app:identifiability}
\textcolor{red}{We show that, given the consistency, strong ignorability, and overlap assumptions, the ITTE 
can be identified. This proof is based on the identifiability proofs in \citet{forastiere2021identification} and \citet{jiang2022estimating}.}

\textcolor{red}{\begin{proof}
From the definition of $\omega_i(t,z)$ and linearity of expectation:
\begin{align}
    \omega_i(t,z) 
&=\mathbb{E}\left[Y_i(t,z) - Y_i(0,0) \mid \mathbf{X_i} = \mathbf{x}_i,\mathbf{X}_{\mathcal{N}_i}=\mathbf{x}_{\mathcal{N}_i}\right], \\
&= \mathbb{E}\left[Y_i(t,z) \mid  \mathbf{X_i} = \mathbf{x}_i,\mathbf{X}_{\mathcal{N}_i}=\mathbf{x}_{\mathcal{N}_i}\right] - \mathbb{E}\left[Y_i(0,0) \mid  \mathbf{X}_i = \mathbf{x}_i,\mathbf{X}_{\mathcal{N}_i}=\mathbf{x}_{\mathcal{N}_i}\right]. \label{eq:cade_decomp}
\end{align}
Using strong ignorability, potential outcomes are independent of treatment and network exposure given $\mathbf{X}_i$ and $\mathbf{X_{\mathcal{N}_i}}$:
\begin{align}
    &\omega_i(t,z)=\mathbb{E}\left[Y_i(t,z) \mid  \mathbf{X}_i=\mathbf{x}_i,\mathbf{X}_{\mathcal{N}_i}=\mathbf{x}_{\mathcal{N}_i},T_i=t,Z_i=z\right] \nonumber \\
&\qquad - \mathbb{E}\left[Y_i(0,0) \mid  \mathbf{X}_i = \mathbf{x}_i,\mathbf{X}_{\mathcal{N}_i}=\mathbf{x}_{\mathcal{N}_i},T_i=0,Z_i=0\right]\label{eq:ignorability_step},
\end{align}
while overlap ensures that the quantities above are well-defined \citep{Hernan2024-HERCIW}.
Finally, by applying consistency, we can identify the ITTE:
\begin{align}
    &\omega_i(t,z)=  \mathbb{E}\left[Y_i \mid  \mathbf{X}_i = \mathbf{x}_i,\mathbf{X}_{\mathcal{N}_i}=\mathbf{x}_{\mathcal{N}_i},T_i=t,Z_i=z\right] \nonumber \\
 & \qquad-\mathbb{E}\left[Y_i \mid  \mathbf{X}_i = \mathbf{x}_i,\mathbf{X}_{\mathcal{N}_i}=\mathbf{x}_{\mathcal{N}_i},T_i=0,Z_i=0\right]\label{eq:consistency_step}.
\end{align}
\end{proof}}

 \textcolor{red}{The proofs for the identifiability of MITE and the spillover effect are analogous.}

\section{\textcolor{red}{Difference between MITE and ITE}}\label{app:MITE v ITE}
Note that the MITE differs from
the traditional ITE as the latter does not take into account graph information and is defined as follows:
\begin{equation}
\tau_{i,trad}=\mathbb{E}\bigl[Y_i(T_i=1)-Y_i(T_i=0)\mid\mathbf{x}_i\bigr].
\end{equation}
The above equation is ill-defined in the presence of interference since $Y_i$ not only depends on $i$'s treatment and features but also on the treatments and features of $i$'s neighbors. We can adapt the traditional definition of the ITE to a network context as follows:
\begin{equation}
\tau_{i,trad}=\mathbb{E}_{(\mathbf{X}_{\mathcal{N}_i},Z_i)}\bigl[\mathbb{E}\left[Y_i(T_i=1,Z_i = z)-Y_i(T_i=0,Z_i=z)\mid\mathbf{x}_i,\mathbf{x}_{\mathcal{N}_i}\right]\bigr].
\end{equation}
In this definition, we marginalize out $\mathbf{X}_{\mathcal{N}_i}$ and $Z_i$. This is what classical causal inference techniques would estimate in the presence of interference. Using the adapted traditional definition of ITE may lead to biased estimates for MITEs (and hence ITTEs) in the presence of interference \citep{forastiere2021identification}. 

\section{\textcolor{red}{Decomposition of the ITTE into spillover effect and MITE}}\label{app:decomposition ITTE}
\textcolor{red}{
The ITTE can be expressed as the sum of the MITE and the spillover effect, as in \Cref{eq:omeqa equivalence}.
\begin{proof}
Recall that the ITTE 
is defined as 
    \begin{equation} \label{eq:itte_ap}
 \omega_i(t,z) = \mathbb{E}\bigl[Y_i(t,z)-Y_i(0,0)\mid\mathbf{x}_i, \mathbf{x}_{\mathcal{N}_i}\bigr].
\end{equation}
If $t=0$, we immediately obtain the definition of the spillover effect:
\begin{equation}
    \omega_i(0,z) = \delta_i(0,z) =\mathbb{E}\bigl[Y_i(0,z)-Y_i(0,0)\mid\mathbf{x}_i, \mathbf{x}_{\mathcal{N}_i}\bigr].
\end{equation}
Which proves the first case of \Cref{eq:omeqa equivalence}.
For $t=1$, using linearity of expectation, we add and subtract $\mathbb{E}\left[Y_i(1,0)\mid\mathbf{x}_i, \mathbf{x}_{\mathcal{N}_i}\right]$ in \Cref{eq:itte_ap}:
\begin{align}
  \omega_i(1,z) &= \mathbb{E}\left[Y_i(1,z) - Y_i(1,0) +Y_i(1,0)-Y_i(0,0)\mid\mathbf{x}_i, \mathbf{x}_{\mathcal{N}_i}\right], \nonumber \\
  &= \delta_i(1,z) + \tau_i(0),
\end{align}
which follows from the definitions of the spillover effect $\delta_i(t,z)$ (\Cref{eq:spillover definition}) and the MITE $\tau_i(z)$ (\Cref{eq:mite}).
Analogously, we add and subtract $\mathbb{E}\bigl[Y_i(0,z)\mid\mathbf{x}_i, \mathbf{x}_{\mathcal{N}_i}\bigr]$:
\begin{align}
  \omega_i(1,z) &= \mathbb{E}\left[Y_i(1,z) - Y_i(0,z) +Y_i(0,z)-Y_i(0,0)\mid\mathbf{x}_i, \mathbf{x}_{\mathcal{N}_i}\right], \nonumber \\
  &= \tau_i(z) + \delta_i(0,z).
\end{align}
\end{proof}
}

\newpage
\section{\textcolor{red}{Pseudocode OTAPI}}\label{app:pseudo}
\begin{algorithm}
\textcolor{red}{
\caption{\textcolor{red}{OTAPI}}
\begin{algorithmic}[1]  
\Require 
  Observational dataset 
    $\mathcal{D}_o = \bigl(\{(\mathbf{x}_i, t_i, y_i)\}_{i=1}^n,\;\mathcal{G}_o\bigr)$; 
  Allocation dataset 
    $\mathcal{D}_a = \bigl(\{\mathbf{x}_j\}_{j=1}^m,\;\mathcal{G}_a\bigr)$;
budget $k$.
\Statex
\State {\bf Step 1: Train relational causal estimator $\mathcal{M}$ on observational data}
\Procedure{TrainModel}{$\mathcal{D}_o$}
    \State Split $\mathcal{D}_o$ into $\mathcal{D}^{train}_o$ and $\mathcal{D}^{val}_o$
    \State Find model $\mathcal{M}^*$ with hyperparameter setting that minimizes validation loss $\mathcal{L}_{val}$
    \State \Return Trained model $\mathcal{M}^*$
\EndProcedure
\Statex
\State {\bf Step 2: Optimize treatment allocation using an optimization algorithm}
\Procedure{AllocateTreatments}
{$\mathcal{M},\mathcal{D}_a,k$}
\State Let $z_j=f(\mathbf{t}_{\mathcal{N}_j})$ be the network exposure mapping for unit $j$
\State The predicted ITTE for unit $j$ is $\hat\omega_j(t_j,z_j)=\mathcal{M}(\mathbf{x}_j,\mathbf{x}_{\mathcal{N}_j},t_j,z_j)$ $-$ $ \mathcal{M}(\mathbf{x}_j,\mathbf{x}_{\mathcal{N}_j},0,0)$
 \State Find treatment allocation $\mathbf{t}^* = (t^*_1, \dots, t^*_m)$ via an optimization algorithm:
        \Statex \hspace{\algorithmicindent} $\mathbf{t}^* \approx \underset{\mathbf{t}}{\arg\max} \sum_{j=1}^m \hat\omega_j(t_j,z_j)$ \Comment{Maximize sum of predicted ITTEs}
        \Statex \hspace{\algorithmicindent} \quad \text{s.t. } $\sum_{j=1}^m t_j \leq k$ \Comment{...within budget $k$ }
    \State \Comment{...using an optimization algorithm (e.g., a genetic or greedy algorithm)}
    \State \Return Treatment allocation vector $\mathbf{t}^*$
\EndProcedure
\Statex
\State $\mathcal{M}^* \gets$ \Call{TrainModel}{$\mathcal{D}_o$}
\State $\mathbf{t}^* \gets$ \Call{AllocateTreatments}{$\mathcal{M}^*, \mathcal{D}_a, k$}
\State \Return  $\mathbf{t}^*$
\end{algorithmic}}
\end{algorithm}

\end{document}